**The effects of diffusional couplings on compositional trajectories and interfacial free energies during phase separation in a quaternary Ni-Al-Cr-Re model superalloy**

Sung-Il Baik[1,2,*], Zugang Mao[1], Qingqiang Ren[1], Fei Xue[1], Carelyn E. Campbell[3], Chuan Zhang[4], Bicheng Zhou[1,5], Ronald D. Noebe[6], David N. Seidman[1,2*]

[1]Department of Materials Science and Engineering, Northwestern University, Evanston, IL 60208, USA.

[2]Northwestern University Center for Atom Probe Tomography (NUCAPT), Evanston, IL 60208, USA

[3.]Materials Science and Engineering Division, National Institute of Standards and Technology (NIST), 100 Bureau Dr. Gaithersburg, MD 20899-8555, USA

[4.] Computherm LLC, 8401 Greenway Blvd. Suite 248, Middleton, WI 53562, USA

[5.] Department of Materials Science and Engineering, University of Virginia, Charlottesville, VA, 22904, USA

[6.] NASA Glenn Research Center, 21000 Brookpark Rd, Cleveland, OH 44135, USA

**Corresponding authors email addresses:* *si-baik@northwestern.edu* (Sung-Il Baik); *d-seidman@northwestern.edu* (David N. Seidman)

***Abstract***

The temporal evolution of ordered γ′($L1_2$)-precipitates and the compositional trajectories during phase-separation of the γ(face-centered-cubic (f.c.c.))- and γ′($L1_2$)-phases are studied in a Ni–0.10Al-0.085Cr-0.02Re (mole-fraction) superalloy, utilizing atom-probe tomography, transmission electron microscopy, and the Philippe-Voorhees (PV) coarsening model. As the γ′($L1_2$)-precipitates grow, the excesses of Ni, Cr and Re, and depletion of Al in the γ(f.c.c.)-matrix develop because of diffusional fluxes crossing the γ(f.c.c.)/γ′($L1_2$) heterophase interfaces. The coupling effects on diffusional fluxes are introduced (PV coarsening model) in terms of the diffusion tensor, **D,** and the second-derivative tensor of the molar Gibbs free energies, $\mathbf{G''}$, obtained employing Thermo-Calc and DICTRA calculations. The Gibbs interfacial free energies, $\sigma^{\gamma/\gamma'}$, are (6.9 ± 1.4) mJ/m$^2$ with all terms in **D** and $\mathbf{G''}$, which changes to (18.9 ± 2.1) mJ/m$^2$, (37.7 ± 3.3) mJ/m$^2$, and (-7.5 ± 1.2) mJ/m$^2$ when not including the off-diagonal terms in **D**, $\mathbf{G''}$, and both **D** and $\mathbf{G''}$, respectively. The experimental APT compositional trajectories are displayed and compared with the PV model in a partial quaternary phase-diagram, employing a partial tetrahedron. The compositional trajectories measured by APT exhibit curvilinear behavior in the nucleation and growth regimes, t < 16 h, which become vectors, moving simultaneously toward the γ(f.c.c.) and γ′($L1_2$) conjugate solvus-surfaces, for the quasi-stationary coarsening regime, t ≥ 16 h. The compositional trajectories for t ≥ 16 h are compared to the PV model with and without the off-diagonal terms in **D** and $\mathbf{G}''$. The directions, including the off-diagonal terms in the **D** and $\mathbf{G}''$ tensors, are consistent with the APT experimental data.

**Nomenclature**

| | |
|---|---|
| $a_\gamma$ | lattice parameter of the γ(f.c.c.)-matrix |
| $a_{\gamma'}$ | lattice parameter of the γ′($L1_2$)-precipitate |
| $C_i$ | concentration of component *i* |
| $C_i^\gamma$ | concentration of component $i$ in the γ(f.c.c.)-phase |
| $C_i^{\gamma'}$ | concentration of component $i$ in the γ′($L1_2$)-phase |
| $C_i^{ff}$ | far-field (ff) matrix concentration of component *i* |
| $C_i^{eq}$ | equilibrium concentration of component *i* |
| $C_i^{\gamma,eq}$ | $C_i^{eq}(\infty)$ in the γ(f.c.c.)-phase |
| $C_i^{\gamma',eq}$ | $C_i^{eq}(\infty)$ in the γ′($L1_2$)-phase |
| $C_v$ | concentration of monovacancies |
| $\mathbf{D}$ | diffusion tensor |
| $\mathbf{D}_\gamma$ | diffusion tensor for the γ(f.c.c.)-matrix |
| $D_{i,j}$ | element of D, for a diffusing species, *i,* with respect to a concentration gradient of species, *j* |
| $D_{i,j}^N$ | $D_{i,j}$ in the number-fixed frame (N) |
| $D_{i,j}^V$ | $D_{i,j}$ in the volume-fixed frame (V) |
| $D_{i,j}^\gamma$ | $D_{i,j}$ in the γ(f.c.c.)-matrix |
| $f$ | fraction of γ′($L1_2$)-precipitates interconnected by necks |
| $\mathbf{G}''$ | second-derivative (Hessian) of the molar Gibbs free energy tensor |
| $\mathbf{G}''_\gamma$ | $\mathbf{G}''$ in the γ(f.c.c.)-matrix |
| $\mathbf{G}''_{\gamma'}$ | $\mathbf{G}''$ in the γ′($L1_2$)-phase |
| $G''_{i,j}$ | element of $\mathbf{G}''$, for *i* with respect to a concentration gradient of component *j* |
| $K$ | rate constant for the coarsening of $\langle R(t)\rangle$ |
| $k_i$ | rate constant of the supersaturation, $\Delta C_i$, for component *i* |
| $n_{at}$ | total number of atoms enclosed within an iso-concentration surface |
| n | one chosen element as the dependent concentration variable in the $n^{th}$ component alloy system. |
| N | number-fixed frame of reference |
| $N_A$ | Avogadro's number |
| $N_i$ | sample size of 3D volumes |
| $N_{ppt}$ | effective number of the γ′($L1_2$)-precipitates measured |
| $N_v(t)$ | number density per unit volume of γ′($L1_2$)-precipitates |
| $\mathbf{M}$ | mobility tensor |
| $\mathbf{M}_\gamma$ | **M** tensor in the γ(f.c.c.)-matrix |
| $M_{k,i}$ | elements of **M**, for a species *k* with respect to the gradient species *i* |
| $r$ | temporal exponent for $\Delta C_i$ according to the PV models |
| $t$ | aging time |
| $t_o$ | time at which quasi-stationary coarsening commences in an alloy |
| $T_t$ | total number of counts for all the species |
| $R$ | radius of a γ′($L1_2$)-precipitate |
| $\langle R(t) \rangle$ | mean radius of γ′($L1_2$)-precipitates |
| $\langle R(t_o) \rangle$ | mean radius at the onset of quasi-stationary coarsening at $t_o$ |
| $R_{ps}$ | radius of a γ′($L1_2$)-precipitate in a projected planar section (PS) |
| $p$ | temporal exponent for $\langle R(t) \rangle$ according to the PV model |
| $q$ | temporal exponent for $N_v(t)$ according to the PV model |
| V | volume fixed frame of reference |
| $V_m^{\gamma'}$ | molar volume of the γ'($L1_2$)- phase |
| γ(f.c.c.) | Face-centered cubic gamma phase |
| γ′($L1_2$) | $L1_2$ ordered gamma prime phase |
| ε | lattice parameter misfit between the γ(f.c.c.)- and γ'($L1_2$)-phases |
| δ | compositional interfacial width between the γ(f.c.c.)- and γ'($L1_2$)-phases |
| $\Delta\bar{\mathbf{C}}^{\gamma\text{-}\gamma'}$ | difference between the equilibrium concentrations of the γ'($L1_2$)- and γ(f.c.c.)-phases |
| $(\Delta\bar{\mathbf{C}}^{\gamma\text{-}\gamma'})^T$ | transpose (T) of $\Delta\bar{\mathbf{C}}^{\gamma-\gamma'}$ |
| $\Delta C_i$ | solute supersaturations for solute component *i* |
| $d\Delta C_i / dt$ | first derivative of the solute supersaturation with respect to time, *t* |
| $\Delta\bar{\mathbf{V}}$ | partial molar volume change |
| $\frac{\partial \mu_i}{\partial C_j}$ | partial derivative of the chemical potential of species *i*, $\mu_i$, with respect to the mole-fraction of species *j* |
| $\frac{\partial G_\gamma^2}{\partial C_i \partial C_j}$ | second-derivative of the molar Gibbs free energy for species *i* with respect to the concentration gradient of species, *j,* in the γ(f.c.c.)-matrix. Simply represented as $\mathbf{G}''_\gamma$ |
| $\dfrac{\partial G_{\gamma'}^2}{\partial C_i \partial C_j}$ | second-derivative of the molar Gibbs free energy for species *i* with respect to the concentration gradient of species, *j,* in the γ′(L1$_2$)-phase. Simply represented as $\mathbf{G}''_{\gamma'}$ |
| $\langle \lambda(t) \rangle$ | mean edge-to-edge distance between neighboring γ′(L1$_2$)-precipitates |
| $\mu_i$ | chemical potential of component, i. |
| $\mu_v$ | chemical potential of a monovacancy. |
| $\eta$ | detection efficiency of the 2-D microchannel plate (MCP), 80% |
| $\rho$ | atomic number density of the γ′(L1$_2$)-phase, 86.22 atoms nm$^{-3}$ |
| 2σ | two standard deviations from the mean |
| $\sigma^{\gamma/\gamma'}$ | interfacial Gibbs free energy between the γ(f.c.c.)- and γ'(L1$_2$)-phases |
| $\phi_{\gamma'}(t)$ | volume fraction of γ'(L1$_2$)-precipitates |
| $\phi_{\gamma'}^{eq}$ | equilibrium volume fraction of γ′(L1$_2$)-precipitates |
| $\omega_t$ | the total number of counts for all species in a precipate |
| APS | advanced photon source at Argonne National Laboratory |
| APT | atom-probe tomography |
| BF TEM | bright-field transmission electron microscopy |
| DF TEM | dark-field transmission electron microscopy |
| DTA | differential thermal analysis |
| EDM | electrical discharge machining |
| KV | Kuehmann and Voorhees |
| LSW | Lifshitz-Slyozov (LS) diffusion-controlled model and the Wagner (W) interface-controlled model |
| MCP | microchannel plate with a detection efficiency of ~80% |
| NLMR method | nonlinear multivariate regression methodology |
| proxigram | proximity histogram |
| NN | Nearest-neighbor distance |
| PV | Philippe-Voorhees coarsening model |
| RM | refractory metal |
| SADP | selected area diffraction pattern |
| TEM | transmission electron microscopy |
| XRD | X-ray diffraction |

**Graphical Abstract**

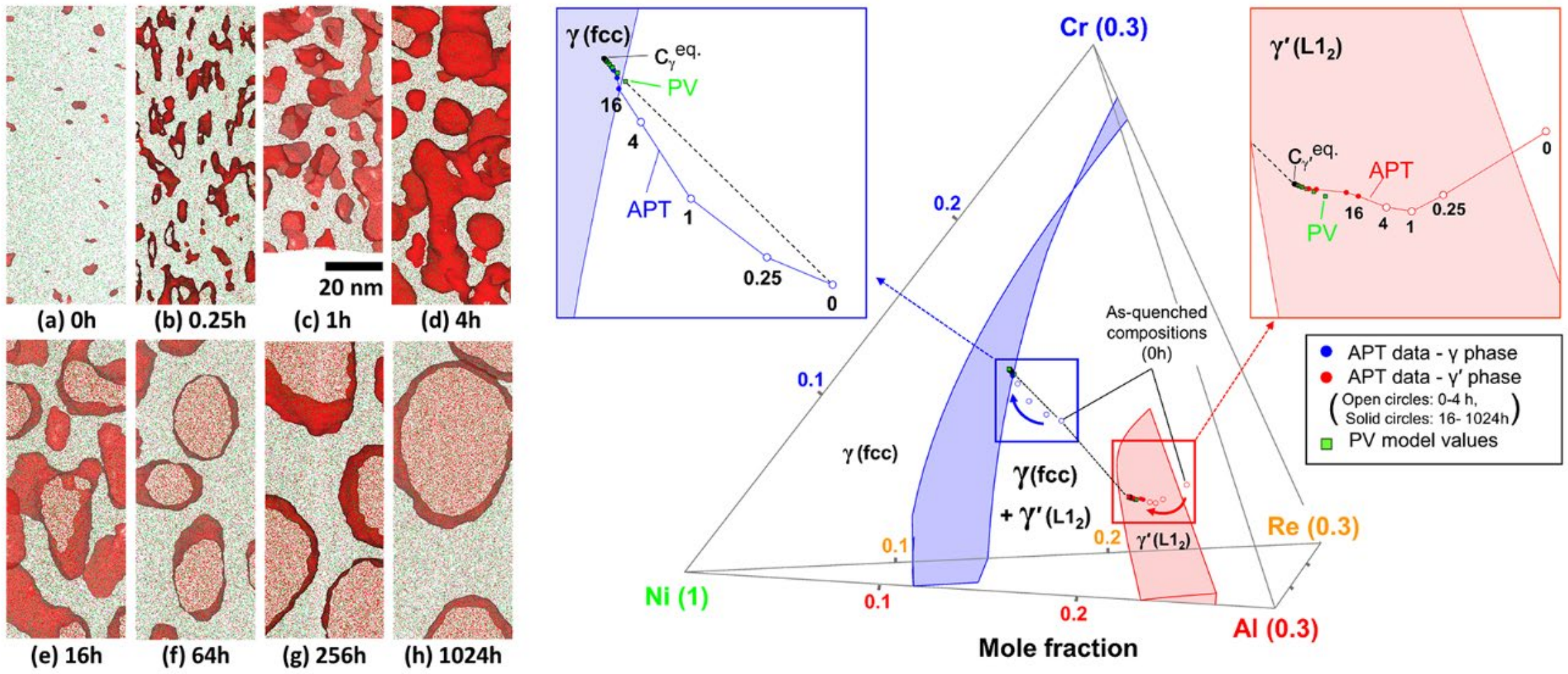

**Figure description**

Temporal evolution of the γ(f.c.c.)- and γ′($L1_2$)-precipitates compositions in a quaternary Ni-0.10Al-0.085Cr-0.02Re (mole fraction) alloy aged at 700 °C (973 K) for times ranging from 0 h through 1024 h (left-hand side) and its three-dimensional (3-D) compositional trajectories in a partial quaternary phase-diagram, utilizing a partial tetrahedron (right-hand side). The compositional trajectories of the γ(f.c.c.)- and γ′($L1_2$)-phases from the atom-probe tomographic (APT) results (blue- and red-circles respectively), and the Phillip-Voorhees (PV) coarsening model (solid-green squares) are successfully captured and compared in the partial tetrahedron (magnified on the left- and right-hand-sides), respectively. The PV model includes coupling of the off-diagonal terms in the Hessian of the Gibbs free energies, $\mathbf{G}''$, and diffusion tensor, $\mathbf{D}$, and the interfacial free energies of the γ(f.c.c.)/γ′($L1_2$) hetero-phase interfaces. The predicted compositional trajectories in the quasi-stationary coarsening regime are in agreement with the APT data.

## 1. Introduction

Nickel-based superalloys are widely used for aircraft jet engines and land-based natural gas combustion turbines for generating electricity because of their superior mechanical properties, and creep and oxidation resistance at elevated temperatures [1-5]. Their excellent mechanical properties at elevated temperatures are derived from the presence of a high-volume fraction of coherent γ′($L1_2$)-precipitates, dispersed in a disordered Ni-rich γ(f.c.c.)-matrix. During the last three decades, the mechanical properties of Ni-based superalloys operating at elevated temperatures have improved significantly, mainly due to the use of high melting-point refractory metal (RM) elements [1, 6]. The superalloys containing RM additions have been the subject of numerous microstructural investigations [7-15]. It is common to use RM elements to decelerate the coarsening kinetics of γ'($L1_2$)-precipitates, while preserving the γ(f.c.c.) plus γ'($L1_2$) microstructure, thereby obtaining potentially longer service times at elevated temperatures. Rhenium, specifically, is of great interest because a small Re-addition improves the creep resistance considerably by increasing the solid-solution strengthening of the γ(f.c.c.)-matrix and decelerating the coarsening kinetics of γ′($L1_2$)-precipitates at elevated temperatures [16-19]. The addition of 0.03 mass fraction Re almost doubles the creep lifetime of single-crystal Ni-based superalloys by decreasing the minimum creep rate and increasing the creep-to-rupture ratio over a wide operating temperature range [16, 20]. Rhenium is enriched at partial dislocations and thereby creates a drag effect on dislocations, which reduces the creep strain rate [21]. It also partitions to the γ(f.c.c.)-matrix and reduces the lattice parameter mismatch between the coherent γ′($L1_2$)-precipitates and the γ(f.c.c.)-matrix, which can decrease diffusional coarsening of γ′($L1_2$) precipitates [22, 23].

The degradation of the stress rupture lifetime in service is dominated by the coarsening behavior of γ′($L1_2$)-precipitates in Ni-based superalloys, which can be predicted by different diffusion-controlled coarsening kinetics models. In a dilute binary alloy, the coarsening of precipitates (Ostwald ripening) is described by the mean-field diffusion-controlled Lifshitz-Slyozov (LS) model [24] or the interface-controlled Wagner model [25]. The diffusion-limited LS model is, however, valid only for a near zero-volume fraction of the second phase because the diffusional interactions, among precipitates are not accounted for. Kuehmann and Voorhees (KV) [26] developed a mean-field model for Ostwald ripening of a ternary alloy by including kinetics and thermodynamic factors. The development of atom-probe tomography (APT) [27-30] enables us to compare and understand the nucleation, growth, and coarsening behavior quantitatively by measuring the compositional trajectories of all the elements, and the temporal development of concentration profiles associated with γ(f.c.c.)/γ′($L1_2$) heterophase interfaces [13, 31-36]. The KV model fails, however, to reproduce the compositional trajectories of the γ′($L1_2$)-precipitates because it does not include coupling terms among the diffusional fluxes, which are responsible for many complex phenomena in multi-component alloys [26, 37]. Later, the Philippe-Voorhees (PV) coarsening model [38] was developed, which is a general coarsening model for a non-dilute multi-component alloy, which accounts for the off-diagonal terms in the diffusion tensor.

In diffusion theory, $\mathbf{D}$ is defined by the product of kinetic and thermodynamic factors employing the following equation [39-41],

$$D_{kj} = \sum_{i=1}^{n} M_{ki} \frac{\partial \mu_i}{\partial C_j} \tag{1}$$

where the $D_{kj}$ is an element of the diffusion tensor, $\mathbf{D}$, for a diffusing species, *k,* with respect to the concentration gradient species, *j*. The $M_{ki}$ are elements of the mobility tensor, $\mathbf{M}$, and the quantity $\frac{\partial \mu_i}{\partial C_j}$ is a thermodynamic factor, which is defined by the partial derivatives of the chemical potentials of species *i*, $\mu_i$, with respect to the mole-fraction of species *j*, $C_j$. This partial derivative of $\mu_i$ is also referred to as second derivatives (Hessian) of the molar Gibbs free energy, $G''_{i,j}$. Whereas $M_{ki}$ is the kinetic factor, which is commonly called the Onsager tensor in the archival literature [42, 43]. Eqn. (1) is expressed in tensorial form by $\mathbf{D} = \mathbf{MG}''$. The $\mathbf{G}''$ and $\mathbf{D}$ are nondiagonal tensors for multi-component alloys because the diffusional fluxes have a multilinear relationship between the sets of the chemical gradients of each specie. The off-diagonal terms in the mobility tensor, $\mathbf{M}$, are related to the frame of reference; zero in the lattice fixed frame of reference and non-zero in the volume-fixed frame of reference due to the existence of the Kirkendall effect [44-46].

The thermodynamic tensor, $\mathbf{G}''$, and mobility tensor, $\mathbf{M}$, for calculating the diffusion tensor, $\mathbf{D}$, are both symmetric and can be obtained employing Thermo-Calc and DICTRA[1] simulation programs [41, 47, 48]. The diffusion model for Ni-Al-Cr-Re alloys is specified by the concentrations of four atomic species, $C_{Ni}$, $C_{Al}$, $C_{Cr}$, $C_{Re}$, and the concentration of monovacancies, $C_v$; these five concentrations sum to unity. The γ′(L1$_2$)-precipitates are coherent with the γ(f.c.c.)-matrix, and the lattice sites are conserved locally during phase separation. For these crystalline phases, diffusion occurs by a vacancy-exchange mechanism; that is, atoms are continuously exchanging places with first nearest-neighbor (NN) vacant lattice sites. DICTRA simulations assume, however, that the vacancy concentration is in local equilibrium, and therefore the gradient of $\mu_v$ is zero, which is not necessarily correct. The number-fixed frame of reference (N) is also adopted with respect to one of the substitutional elements [39, 40, 49]. Assuming that the atomic volumes of the elements are constants, the inter-diffusion matrix in the number-fixed frame of reference is then represented by eliminating the concentration gradient of one chosen component *n*, $D_{i,j}^{N} = D_{i,j}^{V} - D_{i,n}^{V}$, where $D_{i,j}^{V}$ is the diffusivity of component *i* with respect to $C_j$ in the volume fixed-frame of reference (V) [39, 40]. Henceforth, the diffusivity in the number-fixed frame of reference, $D_{i,j}^{N}$, will be denoted by $D_{i,j}$ without including the frame of reference notation. Using this diffusivity expression, the diffusivity of element *n* is eliminated, which results in a $(n-1)$ x $(n-1)$ diffusion matrix, and where the diffusion tensor, $\mathbf{D}$, in the number-fixed frame of reference is asymmetrical. For Ni-Al-Cr-Re alloys, where Ni is treated as an independent component, the inter-diffusivity tensor is given by:

$$\mathbf{D} = \begin{pmatrix} D_{AlAl} & D_{AlCr} & D_{AlRe} \\ D_{CrAl} & D_{CrCr} & D_{CrRe} \\ D_{ReAl} & D_{ReCr} & D_{ReRe} \end{pmatrix} \qquad (2)$$

where the diagonal terms, $D_{AlAl}$, $D_{CrCr}$, and $D_{ReRe}$ denote the direct coefficients, which represent the influence of the concentration gradient of one component on the diffusion rate of the same component. In contrast, the off-diagonal terms are called coupling coefficients, which are responsible for the cross-diffusional effects of each solute element. The off-diagonal terms of the diffusion matrix introduce kinetic couplings among the diffusional fluxes in multi-component alloys [37]. Recently, the off-diagonal terms were shown to affect significantly the evolution of the mean concentrations and the overall precipitation kinetics [50-52]; coupling effects for generalized coarsening kinetics have not yet been studied systematically.

[1] Certain commercial entities, equipment, or materials may be identified in this document to describe an experimental procedure or concept adequately. Such identification is not intended to imply recommendation or endorsement by the National Institute of Standards and Technology, nor is it intended to imply that the entities, materials, or equipment are necessarily the best available for the purpose.

Herein, we present the effects of diffusional coupling on the compositional trajectories of the γ(f.c.c.)- and γ′($L1_2$)-phases during phase-separation and the interfacial free energies, for a quaternary Ni-Al-Cr-Re alloy employing APT, transmission electron microscopy (TEM), and the PV coarsening model. The roles of the off-diagonal diffusion coefficients, $\mathbf{D}$, are utilized in the PV model, which is fully coupled with thermodynamic quantities, $\mathbf{G}''$, and kinetic parameters, **M**, obtained by employing Thermo-Cal and DICTRA simulations. Finally, the effect of diffusional coupling in a diffusion matrix on the temporal evolution of the compositional trajectories is represented utilizing a partial tetrahedron for a partial quaternary phase-diagram for the first time.

## 2. Experimental procedures

The alloy was prepared by induction-melting of relatively high purity elemental constituents under a partial Ar atmosphere and chill-cast in a copper mold to form a polycrystalline master ingot with a target composition of Ni-0.10Al-0.085Cr-0.02Re (mole-fraction[2]). Samples from the master ingot underwent a three-stage heat treatment process: (1) homogenization; (2) a vacancy anneal; and (3) an aging anneal. The cast ingot was fully homogenized in the γ(f.c.c.)-phase field at 1300 °C for 20 h under vacuum and then furnace cooled. The ingot was then sectioned into 1 cm thick slices, which were vacancy annealed at 980 °C for 4 h in a drop-quench furnace and immediately water quenched without exposure to the atmosphere. The vacancy annealing step, just above the γ′($L1_2$) solvus temperature, was used to reduce the concentration of quenched-in vacancies and to suppress γ′($L1_2$)-precipitate formation during the subsequent water-quench. The vacancy annealing temperature is based on differential thermal analyses (DTA) of the solvus temperature, ≈ 922 °C, which was performed on a homogenized sample, at a rate of 10 K $min^{-1}$ in a helium atmosphere, cycled twice through the temperature range of the reaction. Finally, as-quenched samples were aged in the [γ(f.c.c.) plus γ′($L1_2$)]-phase-field at 700 °C, for times ranging from 0.25 h to 1024 h, followed immediately by a quench into ice-brine water. All aging heat treatments were performed in a flowing argon atmosphere to reduce external oxidation.

[2] All other compositions are in mole fractions unless noted

The TEM samples were mounted, ground, and polished to a 1 μm finish, and cut using a rotary disc cutter with a 3 mm diameter. The polished disks were further thinned to TEM transparency using an electrolyte of 10 % by volume of perchloric acid in an ethanol bath at -30 °C. Conventional bright-field (BF), dark-field (DF)-TEM, and selected area diffraction patterns (SADPs) were obtained employing a JEOL ARM 200 TEM operating at 200 kV, utilizing a double-tilt sample holder. The ordered γ′($L1_2$)-precipitates were imaged employing a centered dark-field condition utilizing a $(\bar{1}10)$ superlattice reflection of the ordered $L1_2$-structure of the γ′($L1_2$)-phase. The precipitate radii and areal fractions were determined from TEM micrographs utilizing the program Image-J [53].

APT nanotip specimens were cut using electrical discharge machining (EDM) from each of the aged samples and sharpened by a two-step electro-polishing procedure; 10 % by volume perchloric acid in acetic acid and 2 % by volume perchloric acid in butoxyethanol at (5 to 21) Vdc [54]. The 3-D APT experiments were performed utilizing a local-electrode atom-probe tomograph, LEAP5000XS, which has an 80 % detection efficiency [55]: Cameca Instruments, Inc., Madison, WI. To obtain accurate compositional measurements the experiments were performed using voltage pulses at a pulse-fraction [(pulse-voltage)/(stationary DC voltage)] of 15 %, a pulse repetition rate of 250 kHz, a target detection rate of 0.02 ions pulse$^{-1}$, a specimen temperature of (30.0 ± 0.2) K, and an ambient gauge pressure of < 4.2 x $10^{-9}$ Pa to obtain accurate compositional measurements [56-58]. The recorded 3-D data were analyzed using the program IVAS3.8.2 (Cameca Instruments [59]). The γ(f.c.c.)/γ′($L1_2$) heterophase interfaces were delineated with Al iso-concentration surfaces utilizing the inflection-point technique [60, 61], and compositional information was obtained using the proximity histogram (proxigram) methodology [62]. The standard error of the APT concentrations was calculated based on counting statistics in the 3-D volumes, which is given by $\sigma = \sqrt{C_i(1 - C_i)/\omega_t}$ , where $C_i$ is the concentration of component $i$ and $\omega_t$ is the total number of counts for all the atomic species [63].

## *3. Results*

### 3.1. Atom probe tomography (APT) and transmission electron microscopy (TEM) measurements

The 3-D APT measurements of the γ′(L1$_2$)-precipitate morphologies in the aged Ni-Al-Cr-Re alloy are displayed in **Fig. 1**. Each subset, (40 x 20 x 100) nm$^3$ is part of an analyzed volume that has at least 40 million atoms for adequate counting statistics of the γ′ (L1$_2$)-precipitate. The full 3-D APT reconstructions are provided in the Supplemental Materials A section. The Ni-rich γ (f.c.c.)-matrix appears greenish with the presence of the other elements and the red 0.14 mole-fr. Al iso-concentration surfaces [60] yield a visual comparison of the γ′(L1$_2$)-precipitates as a function of aging time. Very small Al-rich γ′ (L1$_2$)-precipitates, $\langle R \rangle$ = (1.06 ± 0.25) nm, are detected after direct drop-quenching into water after vacancy annealing at 980 °C. This temperature is above the solvus temperature based on DTA measurements at 922 °C and a *Thermo-Calc* assessment at 928 °C using the TCNi8 database [64]. The presence of γ′ (L1$_2$)-precipitates is attributed to the large solute supersaturation in the γ (f.c.c.)-matrix, which forms during quenching from the solution temperature. The γ′(L1$_2$)-precipitates from a supersaturated solid-solution are evolved temporally, to first order, as three processes: nucleation, growth, and coarsening. For the aging times investigated, the nanoscale γ′(L1$_2$)-precipitates are not faceted and do not raft during coarsening, indicating a small lattice parameter misfit between the two-phases with a small interfacial free energy between the γ(f.c.c.)- and γ′(L1$_2$)-phases [65-69]. Quantitative details for the lattice parameter misfit and interfacial free energy are provided later.

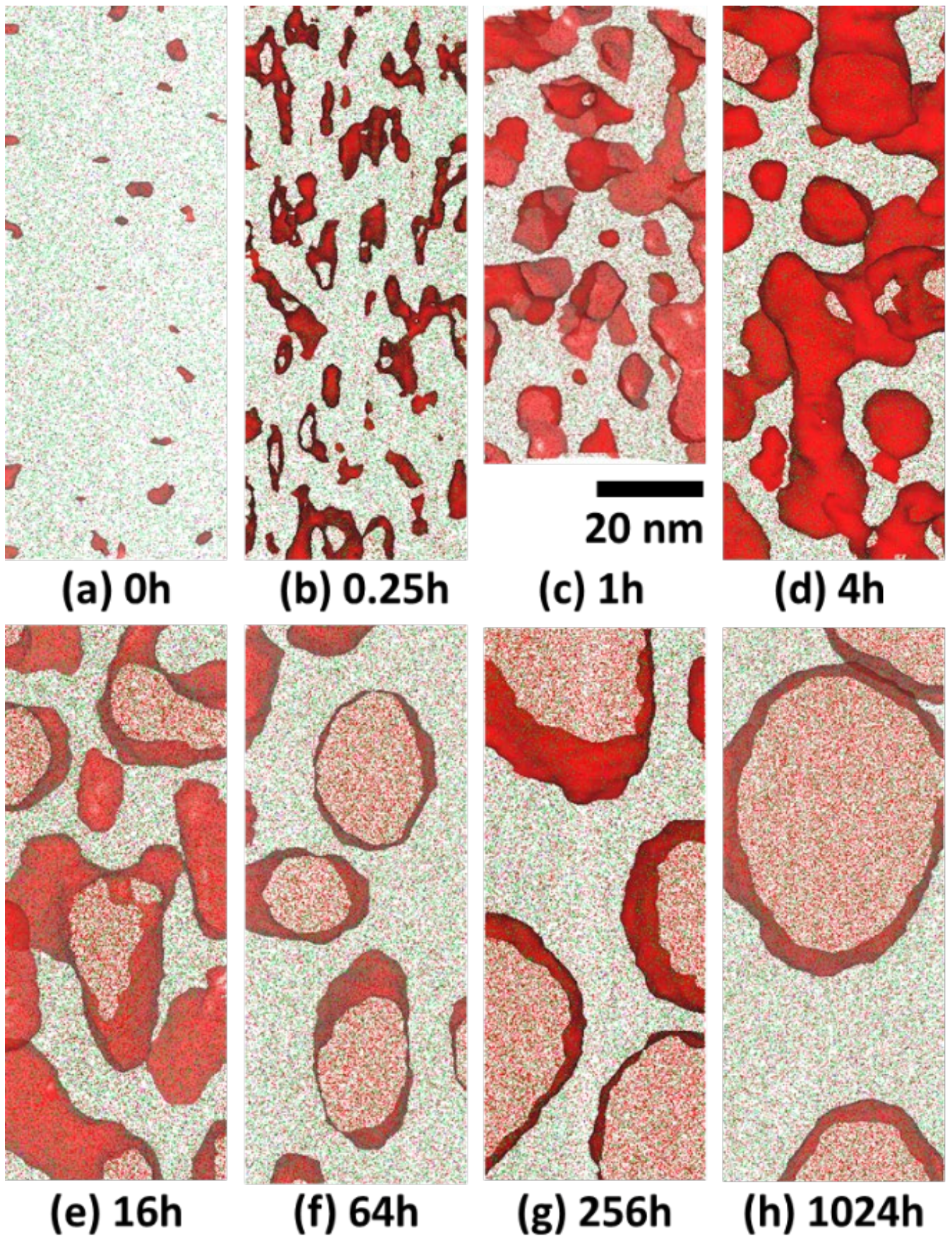

**Fig. 1.** Temporal evolution of γ′(L1$_2$)-precipitates (red iso-concentration surfaces) in the quaternary Ni-0.10Al-0.085Cr-0.02Re alloy aged at 700 °C for times ranging from 0 h through 1024 h. Only a fraction (0.2%) of the Ni (green), Al (red), Cr (blue), and Re (dark yellow) atoms are displayed, for the sake of clarity, and the γ(f.c.c.)/γ′(L1$_2$)-heterophase interfaces are delineated by red 0.14 mole-fr. Al iso-concentration surfaces. Full 3-D reconstructions are provided in the Supplemental Materials A section.

The 3-D APT results are compared to TEM measurements, especially for the coarsening regime where a nearly constant volume fraction of the γ′(L1$_2$)-precipitates is observed. **Fig. 2** displays: (a) a selected area diffraction pattern (SADP); (b-f) dark-field (DF) TEM micrographs of the temporal evolution of the γ′(L1$_2$)-precipitates of the Ni-Al-Cr-Re alloy aged at 700 °C for times ranging from 4 h to 1024 h. The SADP was obtained along a [001] zone-axis, and the superlattice reflections are from the γ′(L1$_2$) ordered-phase within the γ(f.c.c.)-matrix. The dark-field (DF) TEM micrographs were recorded near the [001] zone-axis orientation utilizing two-beam conditions, $\boldsymbol{g}_{\bar{1}10}$. The γ′(L1$_2$)-precipitates are clearly visible with bright-contrast within the black γ(f.c.c.)-matrix background. The γ′(L1$_2$)-precipitates have spheroidal, ellipsoidal or lozenge shaped morphologies, and there is not a preferential alignment of the γ′(L1$_2$)-precipitates along a particular crystallographic direction. The radii of the γ′(L1$_2$)-precipitates were determined from the projected areas of the γ′(L1$_2$)-precipitates measured from the micrographs, assuming circular precipitates on a projected planar section (PS), that is, $R_{ps} = \sqrt{area/\pi}$ . The mean radius of the γ′(L1$_2$)-precipitates is (5.41 ± 1.72) nm at 4 h and they coarsen temporally to (28.76 ± 7.59) nm at 1024 h.

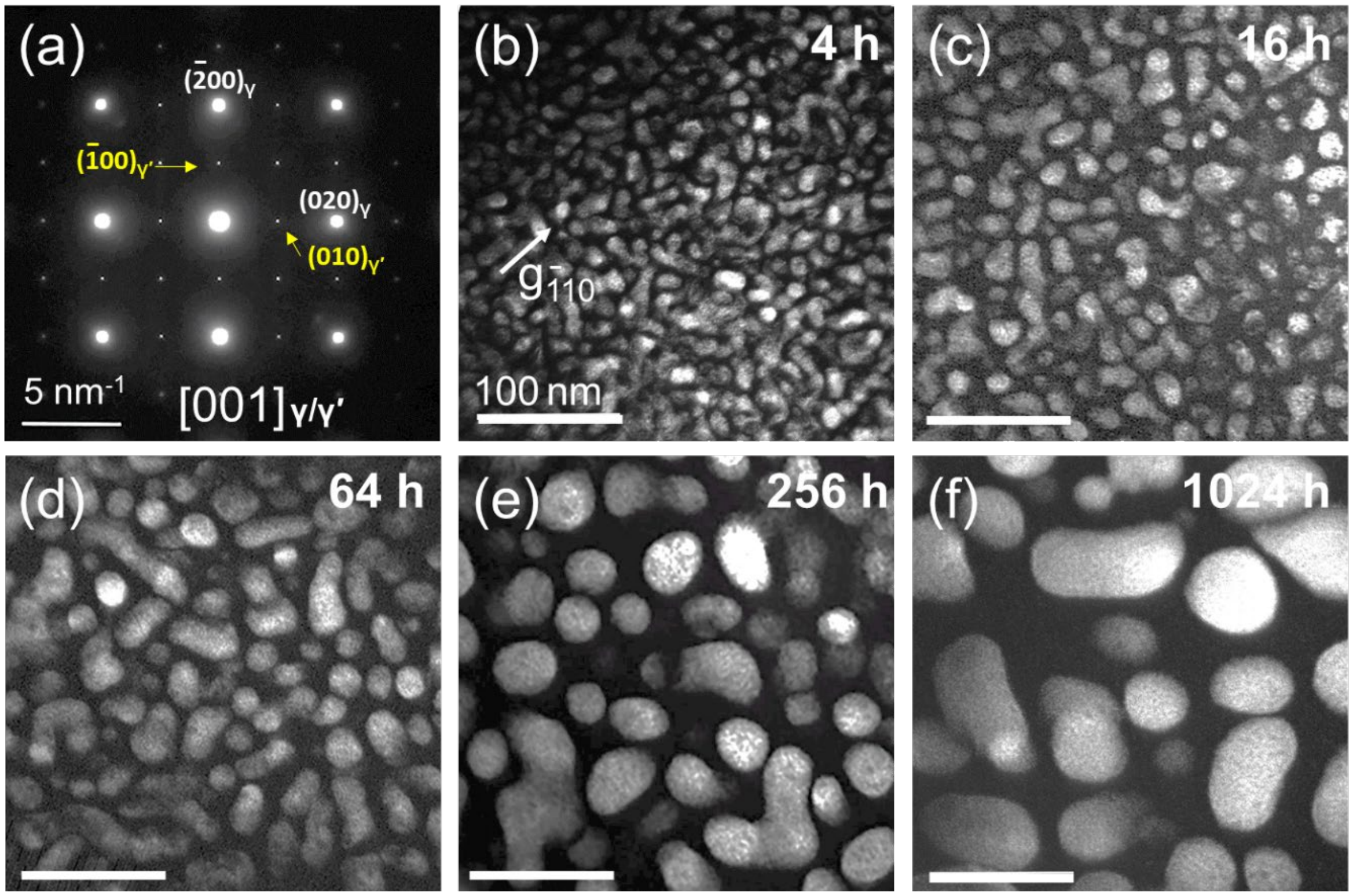

**Fig. 2.** TEM analyses of γ′(L1$_2$)-precipitates in the Ni-0.10Al-0.085Cr-0.02Re alloy: (a) SADP along [100] zone axis (the presence of superlattice reflections in the diffraction patterns are from the γ′(L1$_2$)-precipitates, which reside between the strong reflections from the γ(f.c.c.)-matrix). (b-f). Dark-field (DF) images of the γ′(L1$_2$)-precipitates employing a two-beam condition, $\boldsymbol{g}_{\bar{1}10}$; (b) aged for 4 h; (c) 16 h; (d) 64 h; (e) 256 h; and (f) 1024 h. The lozenge-shaped precipitates are further direct visual evidence for the coagulation-coalescence (C-C) mechanism of coarsening.

### 3.2. Temporal evolutions of γ′(L1$_2$)-precipitates

The temporal evolutions of the γ′(L1$_2$)-precipitates are represented by the mean radius, $\langle R(t)\rangle$, number density, $N_v(t)$, and volume fraction, $\phi_{\gamma'}(t)$, as a function of aging time, *t*, in **Fig. 3**. Their numerical values of are listed in **Table 1**. The error bars represent one standard error of the mean, $\sigma/\sqrt{N_i}$, where $N_i$ is the sample size of 3D volumes. Some error bars are smaller than the marker size. The value of the radius of individual γ′(L1$_2$)-precipitates, *R*, in a 3-D APT reconstruction is determined utilizing the spherical volume equivalent radius method in conjunction with the APT data and the *so-called cluster analysis algorithm* [70] in IVAS,

$$R = \left(\frac{3n_{at}}{4\pi\rho\eta}\right)^{1/3} \tag{3}$$

where $n_{at}$ is the number of atoms enclosed within an iso-concentration surface; $\rho$ is the atomic number density of the γ′(L1$_2$)-phase, 86.22 atoms nm$^{-3}$; and η is the detection efficiency of the 2-D microchannel plate (MCP), 80%, for the LEAP5000XS tomograph. The number density of γ′(L1$_2$)-precipitates, $N_v(t)$, is determined by the number of γ′(L1$_2$)-precipitates contained in the APT reconstructed volumes utilizing a direct counting method [33]. The volume fraction of γ′(L1$_2$)-precipitates, $\phi_{\gamma'}(t)$, is defined by the lever rule from the overall composition of the alloy and the measured compositions of the γ(f.c.c.)-matrix and γ′(L1$_2$)-precipitates at each aging time.

Spheroidal γ′(L1$_2$)-precipitates appear in the as-quenched state, with a mean radius = (1.06 ± 0.25) nm, which then grow and coarsen temporally to $\langle R(t=1024\,h)\rangle$ = (27.17 ± 6.54) nm, which is a factor of 25.6 increase in <R(t)>. These 3-D APT measured values are in excellent agreement with the TEM values, represented as green-open symbols in **Fig. 3(a),** within experimental uncertainty. The $N_v(t)$ values of the γ′(L1$_2$)-precipitates for the aging times investigated is a maximum at 0.25 h [$N_v(t=0.25\,h)$= (83.2 ± 18.0) x 10$^{22}$ m$^{-3}$] and it then decreases continuously with increasing aging time [$N_v(t=1024\,h)$= (0.21 ± 0.06) x 10$^{22}$ m$^{-3}$]. The precipitate volume fraction, $\phi_{\gamma'}(t)$, measured by APT is close to zero at 0 h, [$\boldsymbol{\phi_{\gamma'}(t=0\,h)}$ = (0.38 ± 0.05) %], and is asymptotically approaching a constant value after 16 h, [$\phi_{\gamma'}(t=16h)$= (36.71 ± 3.6) %].

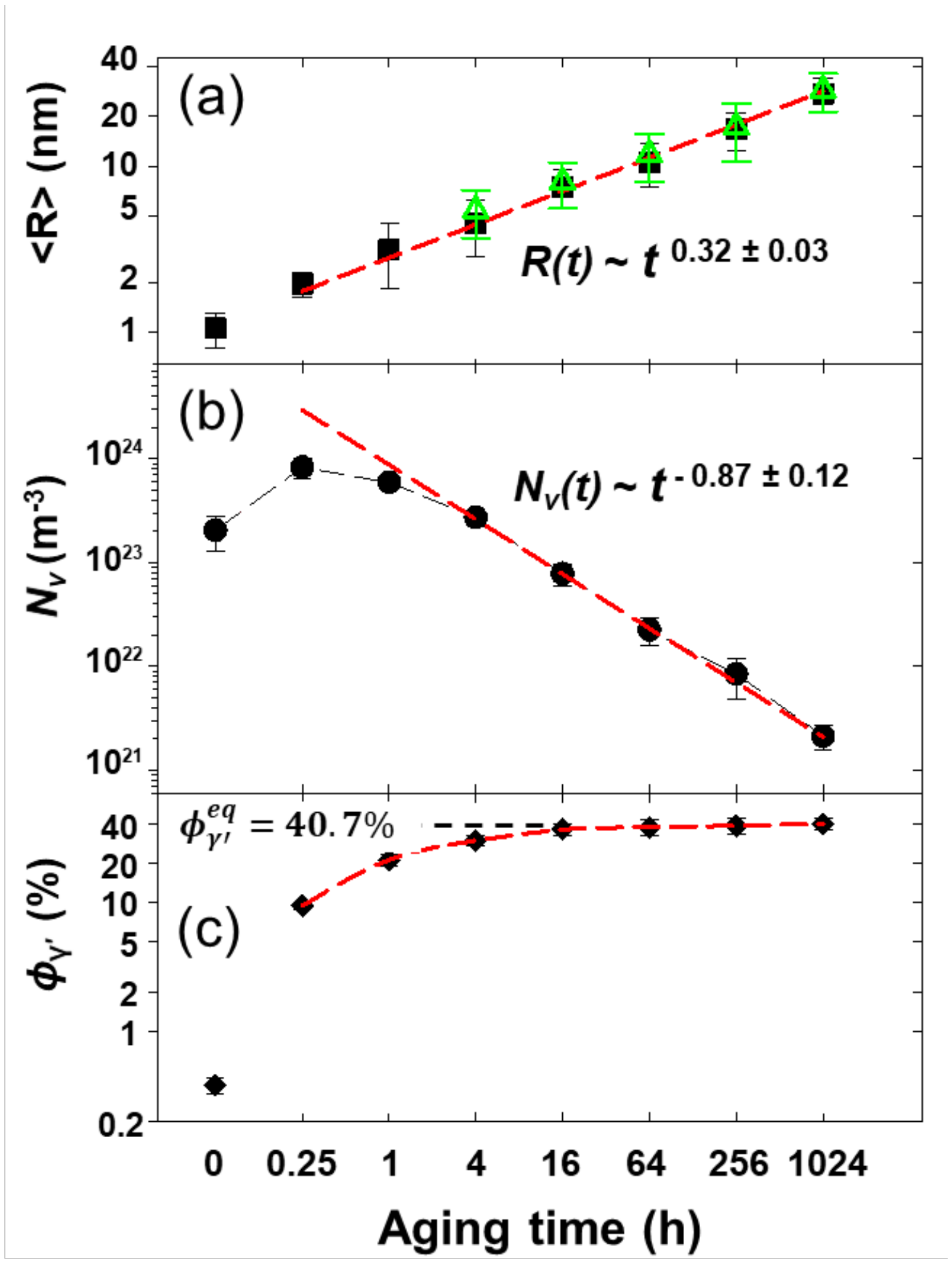

**Fig. 3.** Temporal evolution of the γ′(L1₂)-precipitate: (a) mean radius, $\langle R(t) \rangle$; (b) number density, $N_v(t)$, number per unit volume; and (c) volume fraction, $\phi_{\gamma'}(t)$, of the quaternary Ni-Al-Cr-Ta alloy aged at 700 °C for aging times from 0 h through 1024 h. The TEM measurements of the mean radius, $\langle R(t) \rangle$, are represented by green-colored symbols in (a), which agree with the 3-D APT values, within experimental error. The zero point (0 h) is added as a reference value on the left-hand side of the abscissa, and all values are tabulated in Table 1. The temporal exponents for $\langle R(t) \rangle$ and $N_v(t)$ are calculated to be 0.32 ± 0.03 and -0.87 ± 0.12, respectively, by employing the *nonlinear multivariate regression (NLMR) methodology* [71] to fit the experimental APT and TEM data. The goodness of fit is represented by the minimization of Chi-squared ($\chi^2$), with the details described in the Supplemental Materials C section.

When the γ′($L1_2$)-precipitate phase approaches asymptotically its equilibrium volume fraction (t ≥ 16 h), the alloys enter the diffusion-limited coarsening regime, which is characterized by a steady diminution of $N_v(t)$ and concomitantly increasing values of $\langle R(t)\rangle$. Additonally, the slope of the $N_v(t)$ versus time (t) curve is continuously increasing. Based on the PV model, the rate constants, and the temporal exponents of $\langle R(t)\rangle$ and $N_v(t)$ of the γ′($L1_2$)-precipitates for multi-component alloys are given by the following nonlinear equations:

$$\langle R(t)\rangle^p - \langle R(t_o)\rangle^p = K(t - t_o) \tag{4}$$

$$N_v(t) \cong 0.21 \frac{\phi_{\gamma'}^{eq}}{K} t^{-q} \tag{5}$$

where $K$ is the rate constant for coarsening of $\langle R(t)\rangle$, $\phi_{\gamma'}^{eq}$ is the equilibrium volume fraction of the γ′($L1_2$)-precipitate phase, and $\langle R(t_o)\rangle$ is the mean radius at the onset of quasi-stationary coarsening at time $t_o$. The temporal exponents, $p$ and $q$, are determined employing the NLMR methodology [71] with a minimization of Chi-square ($\chi^2$) to fit the experimental APT data for t ≥ 4 h because the volume fraction change of γ′($L1_2$) is small after 4 h and approaching to the diffusion-limited coarsening regime. The details for these calculations are described in the Supplemental Materials C section. An NLMR analysis of $\langle R(t)\rangle$ yields a coarsening rate constant ($K$) of (6.20 ± 2.40) x $10^{-30}$ $m^3s^{-1}$. The temporal exponent for $<R(t)>$ is $1/p = 0.32 \pm 0.03$; the error in 1/p was calculated using an error propagation analysis. This value matches, within the error, the value p = 3, predicted by the diffusion-limited LS model for binary alloys and PV model for multi-component alloys [38]. Herein, the coarsening rate constant ($K$) is obtained for high equilibrium volume fractions of γ′($L1_2$)-precipitates ($\phi_{\gamma'}^{eq}$ = 0.407). The LS and PV coarsening models in Eqns. 4-5 assume that the precipitate volume fraction is very small, and the diffusion fields of the precipitates do not overlap. This restriction of large volume fractions of γ′($L1_2$)-precipitates in the current study is solved by substituting $f(\phi)K^{PV}$ for $K(\phi)$ with, where $f(\phi)$ is a correction factor that includes the effect of a nonzero volume fraction [72-74]. The theoretical value of ($f(\phi) = 2.45$) in the Marquesee & Ross [75] model is chosen as being representative of high-volume fractions (e.g., $\phi_{\gamma'}^{eq} = 0.407$) and therefore the rate constant for the PV coarsening model ($K^{PV}$) is corrected to be (2.53 ± 0.95) x $10^{-30}$ $m^3s^{-1}$.

The temporal exponent, $q$, of $N_v(t)$ for $t \geq 4$ h yields -0.87 ± 0.12, which is smaller than -1, which is expected for long aging times or higher aging temperatures as described by the PV model [38]. The discrepancy is associated with the high fraction of interconnected necks between adjacent γ′($L1_2$)-precipitates, which is a result of the coagulation-coalescence coarsening mechanism as opposed to the evaporation-condensation mechanism. The value of the pre-factor, $0.21 \times \phi_{\gamma'}^{eq} / K^{PV}$, is (3.38± 0.71) x $10^{28}$ s/m$^3$, which is one order of magnitude larger than the experimental value, (1.10 ± 0.27) x $10^{27}$s/m$^3$.

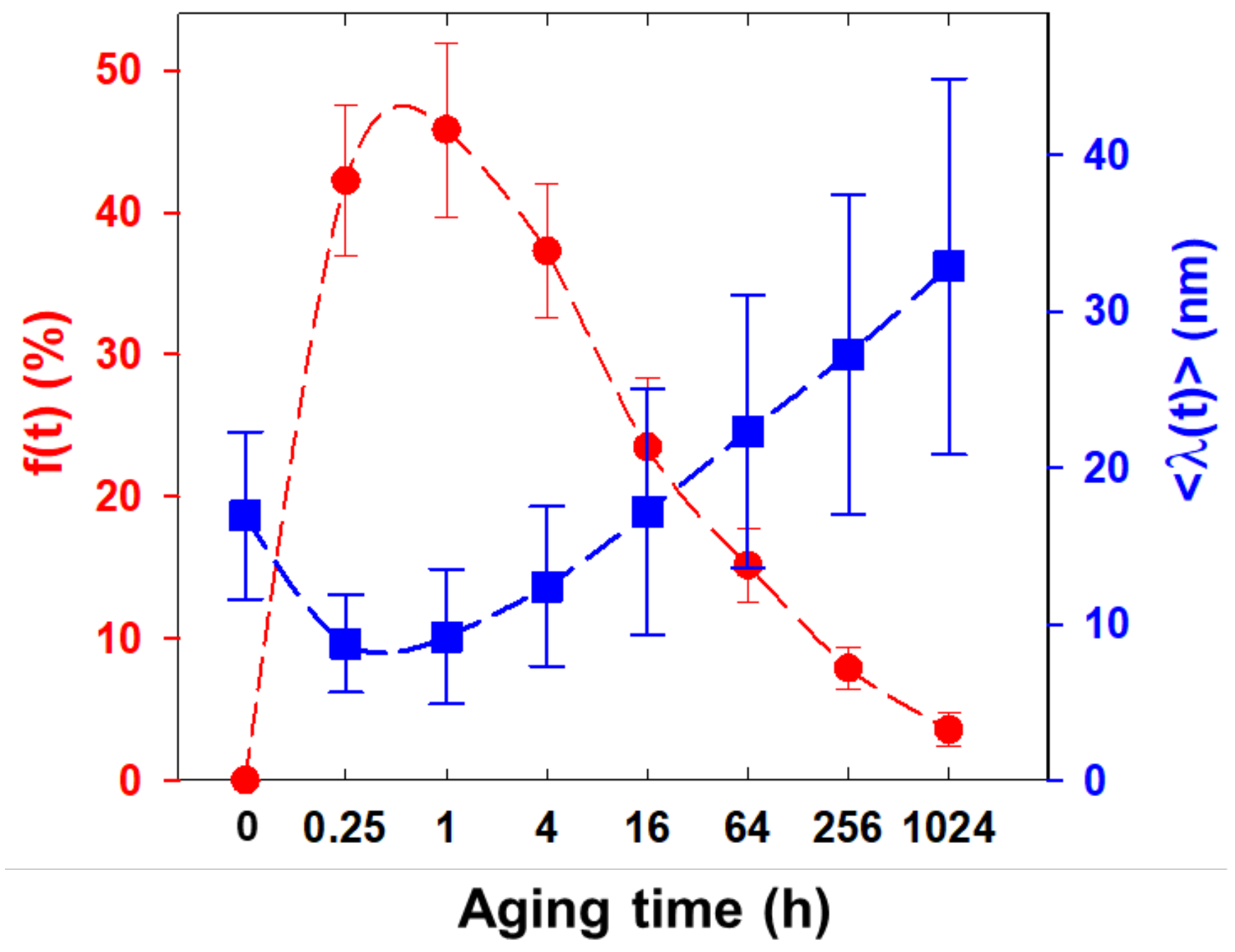

**Fig. 4.** Temporal evolution of the fraction of γ′($L1_2$)-precipitates interconnected by necks, $f(t)$, in percent (%), and the mean minimum edge-to-edge distance between neighboring γ′($L1_2$)-precipitates, $\langle \lambda(t) \rangle$, in nm. A zero point value is added as a reference on the left-hand side of the abscissa and numerical values for both quantities are listed in Table 1. This figure demonstrates clearly that coarsening does not occur through the evaporation-condensation (E-C) mechanism, which is implicit in the Lifshitz-Slyozov (LS) mean-field model of coarsening. The E-C coarsening mechanism is widely assumed to the correct coarsening mechanism in many articles, where the microstructures weren't studied in the same detail as we have performed in this article.

The other important physical result is the high fraction of γ′($L1_2$)-precipitates undergoing coagulation and coalescence (C-C) during the nucleation and growth stage of phase separation. It is clear from the 3-D APT reconstructions that a significant fraction of the γ′($L1_2$)-precipitates are interconnected to one another by necks, which is a direct signature of a coagulation-coalescence mechanism for coarsening [76]: it occurs abundantly in the early stages of phase separation but gradually decreases in the growth and coarsening regimes. **Fig. 4** displays the temporal evolution of the fraction of γ′($L1_2$)-precipitates interconnected by necks, $f(t)$, and the mean edge-to-edge distances, $\langle\lambda(t)\rangle$, between neighboring γ′($L1_2$)-precipitates. The numerical values are listed in Table 1. The mean edge-to-edge distances, $\langle\lambda(t)\rangle$, were measured directly from the x-y-z locations of each γ′($L1_2$)-precipitate in the 3-D APT data, using the Karnesky et al. approach [77], which subtracts the radii of the participating γ′($L1_2$)-precipitates from the distances between the center of each γ′($L1_2$)-precipitate based on the Delaunay triangulation methodology [78]. The value of $\langle\lambda(t)\rangle$ is nearly constant between 0.25 h and 1 h, with values of (8.74 ± 3.11) nm and (9.19 ± 4.32) nm, respectively, and it then increases to (32.84 ± 12.03) nm at 1024 h, as the value of $N_v(t)$ decreases at an approximately constant precipitate volume fraction, $\phi_{\gamma'}(t)$. Whereas the value of $f(t)$ in the 3-D APT measurements display a maximum at 1 h, (45.83 ± 6.12) %, and then it decreases continuously with increasing aging time to (3.57 ± 1.15) % at 1024 h. This decrease in $\langle\lambda(t)\rangle$ corresponds to an increase in $f(t)$ and *vice versa*, as the γ′($L1_2$)-precipitates are close enough to one another for coalescence and coagulation to occur.

For aging times greater than 16 h, the evaporation-condensation mechanism ("the big eat the small" mechanism) becomes dominant over the coagulation-coalescence mechanism, as $\langle\lambda(t)\rangle$ is > 16 nm and $f(t)$ is < 20%. The large values of $\langle\lambda(t)\rangle$ for the γ′($L1_2$)-precipitates limits the formation of ordered necks interconnecting the γ′($L1_2$)-precipitates, by escaping from their overlapping diffusional fields. Historically, it was believed that coarsening of precipitates in dilute solid-solution alloys is accomplished by only the evaporation-condensation mechanism, which is the basis of many mean-field coarsening models [24, 79, 80]. Our present APT results and several prior investigations [13, 31-36] demonstrate that coarsening of γ′($L1_2$)-precipitates, in a concentrated alloy, involves a significant fraction of the γ′($L1_2$)-precipitates undergoing the coagulation-coalescence mechanism, while at longer aging times the evaporation-condensation mechanism dominates where $f(t)$ is approaching zero: at 1024 h $\boldsymbol{f(t)}$ is 3.57 ± 1.15 % and decreasing, while $\langle \lambda(t) \rangle$ is 32.84 ± 12.03 nm and increasing. The large fractions of γ′(L1$_2$)-precipitates undergoing coagulation and coalescence between 0.25 and 1 h, $f(t)$ = (41.67 ± 6.32) and (45.83 ± 6.12) %, respectively, are due to the overlapping behavior of the nonequilibrium concentration profiles surrounding γ′(L1$_2$)-precipitates, which gives rise to diffuse interfaces because of the large value of $N_v(t)$ and concomitantly the small value of $\langle \lambda(t) \rangle$. This is also explained by the diffusional flux-couplings effect among the constituent species toward and away from γ′(L1$_2$)-precipitates [37, 76].

**Table 1.** The temporal evolution of the mean radius, $\langle R(t) \rangle$, number density, $N_v(t)$, volume fraction, $\phi_{\gamma'}(t)$, fraction of γ′(L1$_2$)-precipitates interconnected by necks, $f(t)$, and the edge-to-edge distances between γ′(L1$_2$)-precipitates, $\langle \lambda(t) \rangle$, in the Ni-0.1Al-0.085Cr-0.02Re alloy aged at 700 °C for times ranging from 0 h to 1024 h. The error bars represent one standard error of the mean, $\sigma/\sqrt{N}$.

| Time (h) | $\langle R(t) \rangle$ (nm) | $N_v(t)$ ($10^{22}$ m$^{-3}$) | $\phi_{\gamma'}(t)$ (%) | $f(t)$ (%) | $\langle \lambda(t) \rangle$ (nm) |
|---|---|---|---|---|---|
| 0 | 1.06 ± 0.50 | 20.5 ± 7.48 | 0.38 ± 0.05 | ~0 | 16.93 ± 5.35 |
| 0.25 | 1.96 ± 0.68 | 83.2 ± 18.0 | 9.38 ± 0.71 | 47.92 ± 6.32 | 8.74 ± 3.11 |
| 1 | 3.15 ± 1.68 | 59.2 ± 9.37 | 21.33 ± 2.44 | 45.83 ± 6.12 | 9.19 ± 4.32 |
| 4 | 4.56 ± 3.26 | 27.3 ± 4.72 | 30.21 ± 2.68 | 35.29 ± 4.72 | 12.39 ± 5.09 |
| 16 | 7.13 ± 3.97 | 7.82 ± 0.83 | 36.71 ± 3.60 | 25.47 ± 4.87 | 15.57 ± 6.29 |
| 64 | 10.73 ± 6.15 | 2.25 ± 0.67 | 38.39 ± 5.21 | 19.15 ± 2.56 | 21.31 ± 7.83 |
| 256 | 16.22 ± 8.43 | 0.84 ± 0.15 | 39.31 ± 5.69 | 7.89 ± 1.47 | 27.24 ± 10.23 |
| 1024 | 27.17± 13.72 | 0.21 ± 0.06 | 40.73 ± 4.11 | 3.57 ± 1.15 | 32.84 ± 12.03 |

### 3.3. Compositional evolutions of the γ(f.c.c.)- and γ′($L1_2$)-phases

The concentration profiles straddling the γ(f.c.c.)/γ′($L1_2$) interfaces, for all aging times in the Ni-Cr-Al-Re alloy, are displayed in **Fig. 5**. The compositions of the γ(f.c.c.)-matrix and the γ′($L1_2$)-precipitates evolve temporally as the γ′($L1_2$)-precipitates become enriched in Al and depleted in Ni, Cr, and Re with increasing aging time. Each concentration profile was constructed employing a 0.2 nm bin size with respect to 0.14 mole-fraction of an Al iso-concentration surface, utilizing the proximity histogram (proxigram) methodology [62], from all precipitates in 3D APT reconstructions, see the Supplemental Materials, Figure S1. The zero on the abscissa axis (marked as vertical dotted lines) are placed at the inflection points of the iso-concentration curves, which indicate the locations of the γ(f.c.c.)/γ′($L1_2$)-heterophase interfaces. A negative distance is defined as into the γ(f.c.c.)-matrix, while a positive distance is into the γ′($L1_2$)-precipitates. The error bars represent one standard error in 3-D APT volumes [63]. For the as-quenched state from the solution temperature, the proxigrams display a plateau region in the γ(f.c.c.)-matrix with continuous decreases or increases of each element across the γ(f.c.c.)/γ′($L1_2$)-heterophase interface, spanning the length of the concentration profiles up to ~6 nm into the γ′($L1_2$)-precipitates. The composition of the γ(f.c.c.)-matrix after a solution treatment is 0.7819Ni-0.1051Al-0.0921Cr-0.021Re mole-fr., while the γ′($L1_2$)-nuclei have a solute-supersaturated composition of 0.768Ni-0.1944Al-0.0625Cr-0.0163Re mole-fr. at $<R(t = 0\ h)> = (1.06 \pm 0.25)$ nm.

As the γ′($L1_2$)-precipitates grow, the excesses of Ni, Cr and Re, and depletion of Al in the γ(f.c.c.)-matrix develop because of diffusional fluxes crossing the γ(f.c.c.)/γ′($L1_2$)-heterophase interfaces. These are well represented in Fig. 5 by the evolution of the excesses and depletions in the γ(f.c.c.)-matrix as the γ′($L1_2$) precipitate grows. The mass flux of the excesses and depletions are redistributed to the γ′($L1_2$) precipitates as time increases. The changes of the concentrations can extend beyond the proximity histogram representation. The values of the solute excesses and depletions are represented by solute supersaturations, $\Delta C_i$, which is the difference between the mean and equilibrium concentrations of component *i* at $t = \infty$. The mean composition is measured experimentally from the peak or minimum value of the solute excess or depletion, respectively, because the mean composition of the phase in the far-field (*ff*) is not obtained in an alloy with a high number density of precipitates. The Al-depletion in the γ(f.c.c.)-matrix is large initially and saturates quickly. Whereas Cr and Re are slowly supersaturated in the γ(f.c.c.). For example, Re approaches approximately half of its equilibrium concentration when the Al concentration is close to its equilibrium value. The Ni supersaturation is balanced with respect to the fast-forming Al-depletion and the slow-growing Cr and Re excesses at the γ(f.c.c.)/γ′(L1$_2$)-interface. After aging for 4 h, the change in the solute supersaturations of all the elements is small ($d\Delta C_i / dt \rightarrow 0$), implying the system is in a quasi-stationary coarsening regime as it is slowly approaching its equilibrium volume fraction. The volume-fractions of the γ′(L1$_2$)-precipitates, $\phi_{\gamma'}(t)$, determined by APT measurements, are (30.21 ± 2.68) % at 4 h and (36.71 ± 3.60) % at 16 h, which are somewhat smaller than the equilibrium volume fraction, (40.73 ± 4.11) %, which was determined by extrapolating $\phi_{\gamma'}(t)$ to its equilibrium value, from 4 h to infinite time.

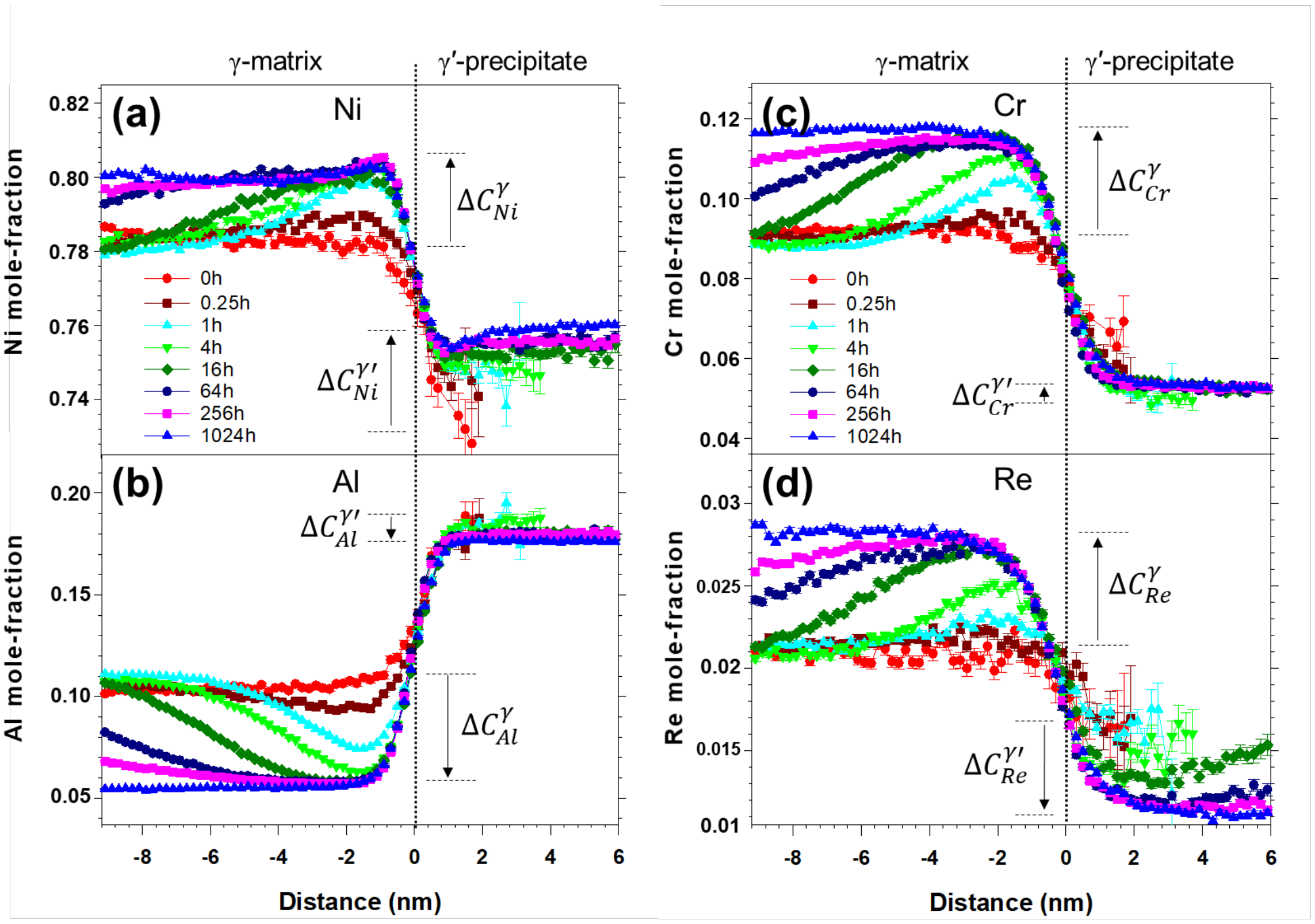

**Fig. 5.** The concentration profiles on either side of the γ(f.c.c.)/γ′(L1$_2$)-heterophase interface of the Ni-Al-Cr-Re alloy aged at 700 °C for aging times from 0 h through 1024 h. The compositions of the two-phases evolve temporally, as the γ(f.c.c.)-matrix becomes enriched in Ni, Cr, and Re, and depleted concomitantly in Al. The dotted-vertical lines are placed at the inflection points of the Al concentration-profiles indicating, by definition, the location of the γ(f.c.c.)/γ′(L1$_2$)-heterophase-interfaces. Note the indicated solute supersaturations, $\Delta C_i$, of the elements Ni, Al, Cr, Re in the

γ(f.c.c.)- and γ′($L1_2$)-precipitate phases. The error bars represent one standard error in the APT measurements. Some error bars are smaller than the marker size.

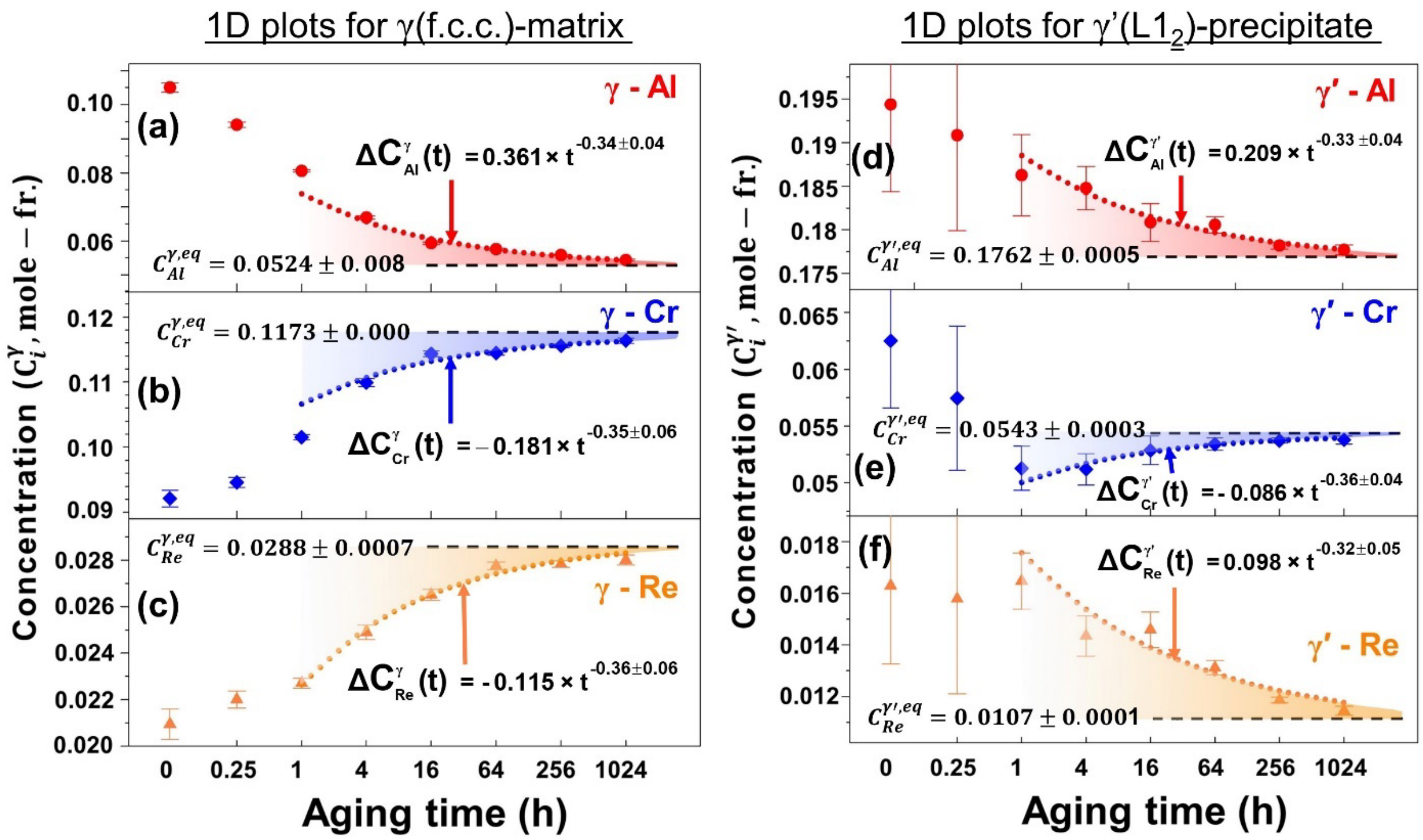

**Fig. 6.** The temporal evolutions of the mean concentrations of Al, Cr, and Re (a, b, c) in the γ(f.c.c.)-matrix, $\boldsymbol{C_i^{\gamma}(t)}$. And the temporal evolution of the Al, Cr, and Re concentrations in the γ′($L1_2$)-precipitates, $C_i^{\gamma'}(t)$, (d, e, f). The six different super- and sub-saturations, $\Delta C_i$, are indicated by shading with three different colors (red, blue and orange), representing the differences between the measured mean concentrations and the equilibrium concentrations (black-dashed horizontal lines with their indicated values). A zero-point (0 h) value is added as a reference state on the left-hand side of each abscissa and their numerical values are listed in Table 2. With increasing aging times, the values of $\Delta C_i$ are continuously decreasing to their equilibrium concentrations, which agrees with the PV model (red, blue, and orange dotted curves). The temporal exponents (*r*) are determined using the NLMR methodology and represented by its reciprocal value (*1/r*). The goodness of fit is represented by the minimization of Chi-squared ($\chi^2$) values, whose details are described in Supplemental Materials Section D.

The temporal evolution of the mean concentration of each element in the γ(f.c.c.) and γ′($L1_2$) phases was derived from proxigram across the γ(f.c.c)/γ′($L1_2$)-heterophase interface, **Fig. 6.** Numerical values are listed in Table 2. The supersaturations, $\Delta C_i$, are indicated by shading with three different colors (red, blue and orange) and the equilibrium concentrations, $C_i^{eq}$, are represented by horizontal black dashed lines. For aging times, $t \leq 1$ h, $\Delta C_i$ is different from the prediction of the PV model because the disparate diffusivities of each element are in the early stages of the nucleation and growth regimes, and the PV model is valid in the coarsening regime, which is when the equilibrium volume fraction of the precipitating phase is finally achieved.

In the stationary coarsening regime, the magnitude of the supersaturation in both the γ(f.c.c) and γ′($L1_2$) phases, $\Delta C_i$, is fit to the following nonlinear equation:

$$\left\langle \Delta C_i(t) \right\rangle = \left\langle C_i^{ff}(t) \right\rangle - \left\langle C_i^{eq} \right\rangle = k_i t^{-1/r} \quad (6)$$

where $k_i$ is the rate constant describing the temporal evolution of the supersaturation, $r$ is the temporal exponent, and $C_i^{eq}$ is an equilibrium concentration. The temporal exponents ($r$) and rate constant ($k_i$) of $\Delta C_i$ were calculated by employing the nonlinear multivariate regression (NLMR) methodology with minimization of the Chi-squared ($\chi^2$) values, which is described in the Supplemental Materials Section C. The equilibrium γ(f.c.c.)-matrix and γ′($L1_2$)-precipitate concentrations, $C_i^{\gamma,eq}$ and $C_i^{\gamma',eq}$, respectively, were obtained by extrapolating the measured concentrations from 4 h to infinite time, yielding 0.8015Ni-0.0524Al-0.1173Cr-0.0288Re (mole-fr.) for the γ(f.c.c.)-matrix, and 0.7588Ni-0.1762Al-0.0543Cr-0.0107 (mole-fr.) for the γ′($L1_2$)-precipitates. The reciprocal of the temporal exponents (*1/r*) of the Al, Cr, and Re supersaturations are 0.34 ± 0.04, 0.35 ± 0.06, 0.36 ± 0.06 for the γ(f.c.c.)-matrix, and 0.33 ± 0.04, 0.36 ± 0.04, 0.32 ± 0.05 for the γ′($L1_2$)-precipitate, respectively. All the temporal exponents agree reasonably well with the LS and PV models value of 3 (1/r = 1/3), within experimental errors. The rate constants for the supersaturations, $k_i$, for Al, Cr, and Re fitted to Eqn. (7) are 0.361 ± 0.16, -0.181 ± 0.13, and -0.115 ± 0.07 mole-fr.$s^{1/r}$ in the γ(f.c.c)-matrix, and 0.209 ± 0.122, -0.086 ± 0.046 and 0.098 ±0.074 mole-fr.$s^{1/r}$ in the γ′($L1_2$)-precipitates, respectively. The effect of the large volume fraction of γ′($L1_2$)-precipitate in the current alloy is also solved by substituting $f(\phi)^{-1/3} k_i^{PV}$ for $k_i$ [34], where $f(\phi)$ is 2.45 at the equilibrium volume fraction ($\phi_{\gamma'}^{eq}$ = 0.407), which is based on the Marquesee & Ross model [75]. The corrected rate constants in the equation for the supersaturations, $k_i^{PV}$ , for Al, Cr, and Re are 0.487 ± 0.216, -0.244 ± 0.175, and -0.115 ± 0.094 mole-fr.s$^{1/r}$ in the γ(f.c.c)-matrix, and 0.282 ± 0.164, -0.116 ± 0.062 and 0.132 ±0.102 mole-fr.s$^{1/r}$ in the γ′(L1$_2$)-precipitates, respectively. For these values, the six different supersaturations decay to zero with different amplitudes of the rate constants ($k_i$). The direction of the supersaturation is represented by positive or negative values associated with the rate constants: a positive value of $k_i$ for Al in the γ(f.c.c)-matrix; and also for Al and Re in the γ′(L1$_2$)-precipitates, implies that the measured concentration is greater than the equilibrium concentration. A negative value means the concentration is approaching its equilibrium from below the equilibrium concentration, **Fig 6**. These signs are related to the thermodynamic factors, $\mathbf{G}''$, in the PV coarsening model (Eqns. 11 and 12 below), which govern the directions of the compositional trajectories in a partial quaternary phase diagram as represented in a partial tetrahedron.

**Table 2.** Temporal evolution of the concentrations in the γ(f.c.c)-matrix and the γ′(L1$_2$)-precipitate phase extrapolated to t = ∞ to obtain the equilibrium concentrations at 700 °C (973 K). The large scatter in the values of the concentrations for aging times, $t \leq 1$ h, are due to the disparate diffusivities of each element in the early stages of the nucleation and growth regimes, which is the transient regime. The error bars represent one standard error in APT measurements.

| 973 K | γ(f.c.c)-Matrix composition (mole-fr.) | | | |
|---|---|---|---|---|
| Time (h) | $C_{Ni}^{\gamma}$ | $C_{Al}^{\gamma}$ | $C_{Cr}^{\gamma}$ | $C_{Re}^{\gamma}$ |
| 0 | 0.7819 ± 0.0019 | 0.1051 ± 0.0014 | 0.0921 ± 0.0013 | 0.0210 ± 0.0006 |
| 0.25 | 0.7893 ± 0.0010 | 0.0941 ± 0.0008 | 0.0946 ± 0.0008 | 0.0220 ± 0.0004 |
| 1 | 0.7952 ± 0.0006 | 0.0806 ± 0.0004 | 0.1015 ± 0.0004 | 0.0227 ± 0.0002 |
| 4 | 0.7983 ± 0.0008 | 0.0668 ± 0.0005 | 0.1099 ± 0.0006 | 0.0249 ± 0.0003 |
| 16 | 0.7998 ± 0.0006 | 0.0594 ± 0.0003 | 0.1143 ± 0.0005 | 0.0265 ± 0.0002 |
| 64 | 0.8002 ± 0.0004 | 0.0576 ± 0.0002 | 0.1144 ± 0.0003 | 0.0277 ± 0.0002 |
| 256 | 0.8008 ± 0.0003 | 0.0559 ± 0.0002 | 0.1155 ± 0.0003 | 0.0278 ± 0.0001 |
| 1024 | 0.8013 ± 0.0005 | 0.0544 ± 0.0003 | 0.1164 ± 0.0004 | 0.0280 ± 0.0002 |
| eq. comp | 0.8015 ± 0.0010 | 0.0524 ± 0.0008 | 0.1173 ± 0.0009 | 0.0288 ± 0.0007 |

| | γ'(L1$_2$)-precipitate composition (mole-fr.) | | | |
|---|---|---|---|---|
| Time (h) | $C_{Ni}^{\gamma'}$ | $C_{Al}^{\gamma'}$ | $C_{Cr}^{\gamma'}$ | $C_{Re}^{\gamma'}$ |
| 0 | 0.7268 ± 0.0085 | 0.1944 ±0.0100 | 0.0625 ± 0.0059 | 0.0163 ± 0.0030 |
| 0.25 | 0.7359 ± 0.0122 | 0.1908 ±0.0109 | 0.0574 ± 0.0064 | 0.0158 ± 0.0037 |
| 1 | 0.7460 ± 0.0048 | 0.1863± 0.0043 | 0.0513 ± 0.0024 | 0.0165 ± 0.0014 |
| 4 | 0.7497 ± 0.0037 | 0.1848± 0.0033 | 0.0512 ± 0.0019 | 0.0143 ± 0.0011 |
| 16 | 0.7517 ± 0.0025 | 0.1809 ± 0.0022 | 0.0528 ± 0.0013 | 0.0146 ± 0.0007 |
| 64 | 0.7529 ± 0.0011 | 0.1806 ± 0.0009 | 0.0534 ± 0.0006 | 0.0131 ± 0.0003 |
| 256 | 0.7562 ± 0.0006 | 0.1782 ± 0.0004 | 0.0537 ± 0.0002 | 0.0119 ± 0.0001 |
| 1024 | 0.7571 ± 0.0009 | 0.1777 ± 0.0006 | 0.0538 ± 0.0004 | 0.0114 ± 0.0002 |
| eq. comp | 0.7588 ± 0.0006 | 0.1762 ± 0.0005 | 0.0543 ± 0.0003 | 0.0107 ± 0.0001 |

## 4. Discussion

### *4.1. Diffusional coupling effects on the interfacial free energies, $\sigma^{\gamma/\gamma'}$.*

The interfacial Gibbs free energy, $\sigma^{\gamma/\gamma'}$, for γ(f.c.c.)/γ′(L1$_2$)-heterophase interfaces is related to the microstructural stability of many precipitation-strengthened alloys and is utilized to predict and improve the creep strength of existing superalloys. The driving force for coarsening is the minimization of the overall Gibbs interfacial free energy of a system subject to the condition that the volume fraction of the precipitating phase has achieved its equilibrium value [24-26, 80, 81]. The interfacial Gibbs free energies were obtained employing experimental coarsening data based on the PV coarsening model [38, 82].

The coarsening rate constant, *K*, for the temporal evolution of the mean precipitate radius, <*R*(t)>, measured utilizing APT and TEM experimental results is employed to calculate the interfacial Gibbs free energy ($\sigma^{\gamma/\gamma'}$) of the γ(f.c.c.)/γ′(L1$_2$)-heterophase interfaces. The PV model assumes non-ideal and non-dilute solid-solutions and includes the off-diagonal terms in the diffusivity and mobility tensors. The interfacial Gibbs free energy, $\sigma^{\gamma/\gamma'}$, is determined from [38]:

$$\sigma^{\gamma/\gamma'} = \frac{9K^{PV}(\mathbf{\Delta\overline{C}}^{\gamma\text{-}\gamma'})^T \mathbf{M}_{\gamma}^{-1}\mathbf{\Delta\overline{C}}^{\gamma\text{-}\gamma'}}{8V_m^{\gamma'}} \quad (7)$$

where $V_m^{\gamma'}$ is the molar volume of the γ′(L1₂)-phase, $\mathbf{\Delta \bar{C}}^{\gamma\text{-}\gamma'}$ is the difference between the equilibrium concentrations of the γ'(L1₂)- and γ(f.c.c.)-phases, and the superscript letter T indicates its transpose. The quantity $V_m^{\gamma'}$ is defined to be $N_A(a_{\gamma'})^3/4$, where $N_A$ is Avogadro's number, and $a_{\gamma'}$ is the lattice parameter of the γ′(L1₂)-phase. The lattice parameter was measured using a synchrotron XRD pattern at 700 °C, **Supplementary Material Section B**: its value is 0.3591 nm, which yields $V_m^{\gamma'} = 6.948 \times 10^{-6}$ m$^3$/mol. As embodied in Eqn. (1), the mobility tensor in the γ(f.c.c.)-matrix, $\mathbf{M}_\gamma$, combined with a thermodynamic factor yields diffusivities through the relationship, $\mathbf{D}_\gamma = \mathbf{M}_\gamma \mathbf{G}''_\gamma$, where $\mathbf{G}''_\gamma$ is the Hessian: that is, the second-derivatives of the Gibbs free energies in the γ(f.c.c.)-phase at its equilibrium concentrations, $C_i^{\gamma,eq}$.

The nominal composition of the alloy measured by the APT, 0.785Ni-0.103Al-0.091Cr-0.021Re mole-fr., is taken as a reference state for the Thermo-Calc and DICTRA computations. Thermodynamic equilibrium, given by the common tangent line to the Gibbs free energy curves of the two phases [γ(f.c.c.) and γ′(L1₂)], yields 0.806Ni-0.052Al-0.110Cr-0.032Re mole-fr. for the γ(f.c.c.)-phase and 0.761Ni-0.152Al-0.076Cr-0.011Re mole-fr. for the γ′(L1₂)-phase. The thermodynamic calculations predict partitioning of Al to the γ′(L1₂)-phase and partitioning of Ni, Cr, and Re to the γ(f.c.c.)-phase, which is in good agreement with our APT experiments, **Fig.4**. Using Ni as the reference component for the $\mathbf{G}''_\gamma$ and $\mathbf{D}_\gamma$ tensors, of the γ(f.c.c.)-phase, they were computed utilizing the thermodynamic database TCNI8 and the Ni-mobility database [41, 64]:

$$\mathbf{G}''_\gamma = \begin{pmatrix} \frac{\partial G_\gamma^2}{\partial^2 C_{Al}} & \frac{\partial G_\gamma^2}{\partial C_{Al}\partial C_{Cr}} & \frac{\partial G_\gamma^2}{\partial C_{Al}\partial C_{\mathrm{Re}}} \\ \frac{\partial G_\gamma^2}{\partial C_{Cr}\partial C_{Al}} & \frac{\partial G_\gamma^2}{\partial^2 C_{Cr}} & \frac{\partial G_\gamma^2}{\partial C_{Cr}\partial C_{\mathrm{Re}}} \\ \frac{\partial G_\gamma^2}{\partial C_{\mathrm{Re}}\partial C_{Al}} & \frac{\partial G_\gamma^2}{\partial C_{\mathrm{Re}}\partial C_{Cr}} & \frac{\partial G_\gamma^2}{\partial^2 C_{\mathrm{Re}}} \end{pmatrix} = \begin{pmatrix} 3.31\times10^5 & 1.34\times10^5 & 1.48\times10^5 \\ 1.20\times10^5 & 1.59\times10^5 & 0.12\times10^5 \\ 1.30\times10^5 & 0.08\times10^5 & 3.56\times10^5 \end{pmatrix} \mathrm{J{\cdot}mol^{-1}} \quad (8)$$

and

$$\mathbf{D}_\gamma = \begin{pmatrix} D^\gamma_{AlAl} & D^\gamma_{AlCr} & D^\gamma_{AlRe} \\ D^\gamma_{CrAl} & D^\gamma_{CrCr} & D^\gamma_{CrRe} \\ D^\gamma_{ReAl} & D^\gamma_{ReCr} & D^\gamma_{ReRe} \end{pmatrix} = \begin{pmatrix} 1.17\times10^{-18} & 4.68\times10^{-19} & 5.31\times10^{-19} \\ 1.35\times10^{-19} & 2.65\times10^{-19} & -9.38\times10^{-21} \\ -2.84\times10^{-20} & -1.54\times10^{-20} & -4.49\times10^{-21} \end{pmatrix} \text{ m}^2\text{s}^{-1} \quad (9)$$

The reduction of the number of equations from *n* to *n-1* is obtained from the definition of the second derivatives of the Gibbs free energies, $\mathbf{G}''_\gamma$, and the diffusivities $\mathbf{D}_\gamma$, with one chosen element (Ni) as the reference component utilizing a number-fixed frame of reference (N) [39, 40, 49]. The positive values of the inter-diffusivities in $\mathbf{D}_\gamma$ indicate that diffusional flux is causing phase separation, whereas the negative values indicate that the diffusional flux is from the γ′($L1_2$)-precipitates to the disordered γ(f.c.c.)-phase.

Based on our experimental APT and TEM data, the coarsening rate constant associated with the mean radius, $K^{PV}$, was calculated, using **Fig. 2** and Eqn. 4, to be (2.53 ± 0.95) x $10^{-30}$ $m^3s^{-1}$, for the quaternary Ni-Al-Cr-Re alloy. Employing the $K^{PV}$ value, the interfacial free energy, $\sigma^{\gamma/\gamma'}$, is calculated, using Eqns. (7-9), to be: (i) (6.9 ± 1.4) mJ/$m^2$ at 700 ºC. The calculated value of $\sigma^{\gamma/\gamma'}$ is strongly influenced by the off-diagonal terms in the $\mathbf{D}$ and $\mathbf{G}''$ tensors, which are included in the PV model. The Gibbs interfacial free energy, $\sigma^{\gamma/\gamma'}$, changes to: (ii) (18.9 ± 2.1) mJ/$m^2$ *when not including the off-diagonal terms in* $\mathbf{D}$; (iii) (37.7 ± 3.3) mJ/$m^2$ *without the off-diagonal terms in* $\mathbf{G}''$; and (iv) (-7.5 ± 1.2) mJ/$m^2$ *without the off-diagonal terms in both the* $\mathbf{D}$ *and* $\mathbf{G}''$ *tensors.* These differences are important because it demonstrates how the coupling behavior of solute elements affects ultimately the coarsening behavior of precipitates in alloys. In general, a higher Gibbs interfacial free energy induces a faster coarsening rate. The negative value of the Gibbs interfacial free energy is nonphysical because it causes the overall free energy of a system to decrease continuously. J. W. Gibbs (1876) stated that a system would continue making interfaces in a system with a negative interfacial free energy until classical thermodynamics is no longer applicable [83]. A similiar point was made by Cahn and Hilliard with respect to segregation at grain boundaries [84].

### *4.2 The diffusional coupling effects on compositional trajectories*

The Gibbs-Thompson effect describes the compositional change of a precipitate or matrix due to a change of the bulk free energy caused by the curvature of an interfacial surface under tension, which yields the solute supersaturation at a matrix/precipitate heterophase-interface as a function of the mean curvature of a precipitate [81, 85, 86]. In this section, the compositional trajectories of the γ(f.c.c.)- and γ′($L1_2$)-phases during stationary coarsening are compared with the predictions of the PV model [38].

The concentration profiles and the supersaturations of each element in the Ni-Al-Cr-Re quaternary alloy, **Figs. (5, 6)**, demonstrate that the compositions of the γ(f.c.c.)- and γ′($L1_2$) -phases are initially highly supersaturated and evolve temporally toward equilibrium compositions with increasing aging time. By employing the concentrations of the four elements, the compositional trajectories of the γ(f.c.c.)-matrix and γ′($L1_2$)-precipitate phases can be represented in a partial Ni-Al-Cr-Re quaternary phase diagram, Fig. 7, employing a partial tetrahedron. The light-blue- and light-red-colored surfaces represent calculated conjugate solvus surfaces of the γ(f.c.c.)- and γ′($L1_2$)-phases, respectively, utilizing *CompuTherm with the Pandat-Nickel database* [87]. The initial composition of the alloy is the as-quenched composition at t = 0, which lies in the [γ(f.c.c.) plus γ′($L1_2$)] phase-field between the two conjugate solvus surfaces. The APT experimental data points denoted by blue- and red-circles represent the compositional trajectories of the γ(f.c.c.)- and γ′($L1_2$)-phases, respectively. The highly magnified regions to the left- and right-hand sides of the partial tetrahedron display the compositional trajectories of both phases in detail.

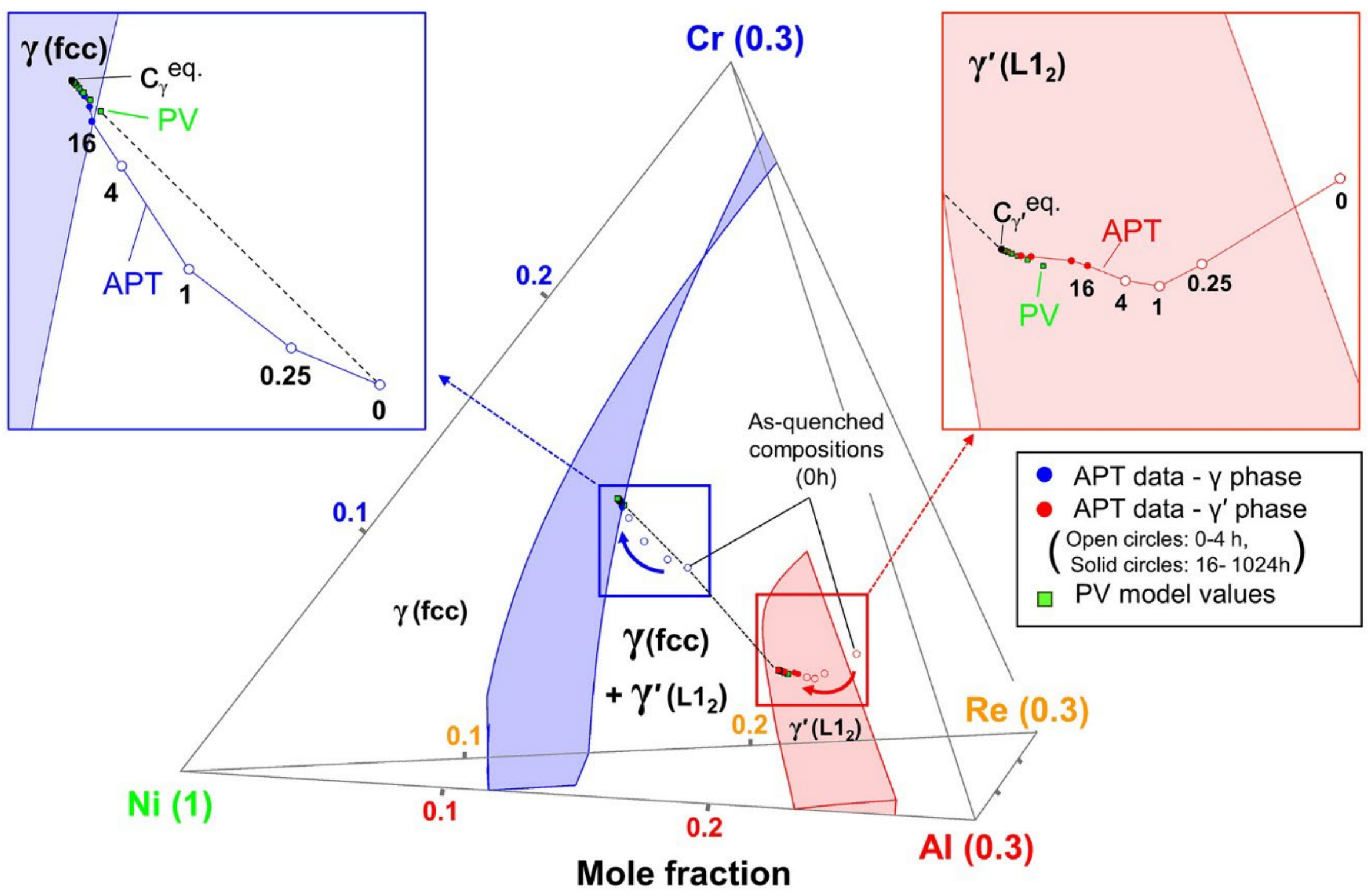

**Fig. 7.** A graphical representation of the composition trajectories for the quaternary Ni-Al-Cr-Re alloy, employing a partial tetrahedron, to display the experimental (APT) data and calculated values, employing the PV coarsening model. The equilibrium tie-line connecting the equilibrium compositions between the γ(f.c.c.)- and γ′(L1$_2$)-phases, at infinite time, is indicated by a black dashed-line, which is a vector. All the experimental data points (APT) are indicated by the blue- and red-circles for the γ(f.c.c.)- and γ′(L1$_2$)-phases, respectively, which commence with their initial compositions at $t = 0$ h and terminate at their final equilibrium compositions on the two-conjugate solvus-surfaces. The open-blue- and open-red-circles are for $t \leq 4$ h, whereas the solid-blue and solid-red-circles are for APT data in the quasi-stationary coarsening regime, $t \geq 16$ h. The compositions calculated from the PV model are represented by solid-green squares. The behaviors of the supersaturations in the γ(f.c.c.)-matrix and γ′(L1$_2$)-precipitates are magnified on the left- and right-hand-sides of the partial tetrahedron, respectively, for the sake of clarity.

The compositional trajectory of the γ(f.c.c.)-phase is initially curvilinear because the volume fractions have not yet achieved their quasi-stationary values; the nucleation and growth regimes are dominant between 0 h and 4 h aging times, and the compositions are most affected by the elements with the largest diffusivities, particularly Al. The APT experimental trajectories are represented as open-blue-circles to distinguish them from the quasi-stationary coarsening regime for t ≥ 16 h, which are represented by solid-blue circles. The compositional trajectories in the coarsening regime become linear, a vector, terminating on the light-blue conjugate solvus surface, whose final composition is the equilibrium concentration of the γ(f.c.c.)-phase. The γ′($L1_2$)-compositional trajectory, on the right-hand side of the γ′($L1_2$)-conjugate solvus surface (light-red color), commences at 0 h deep inside the γ′($L1_2$)-phase-field. At t = 0 h, the γ′($L1_2$)-phase has already been nucleated with $\langle R(t=0)\rangle$ equal to (1.06 ± 0.25) nm, and $\phi_{\gamma'}(t)$ equal to (0.38 ± 0.05) %, which is due to the large solute supersaturation in the γ(f.c.c.)-matrix during drop-quenching after vacancy annealing at 980 ºC. This compositional trajectory is also initially curvilinear, and it becomes a vector as it moves toward the light-red conjugate solvus surface.

The curvilinear behavior of the compositional trajectories is due to the different diffusivities of solute elements during the initial stages of phase separation; Al diffuses significantly faster than Cr and Re. The diagonal diffusivity of Al in the γ(f.c.c.)-matrix, $D^{\gamma}_{AlAl} = 1.17\times10^{-18}$ $m^2sec^{-1}$, is one or two orders of magnitude greater than the diffusivities of Cr and Re, $D^{\gamma}_{CrCr} = 2.65\times10^{-19}$ $m^2sec^{-1}$ and $D^{\gamma}_{\mathrm{ReRe}} = 4.49\times10^{-21}$ $m^2sec^{-1}$, respectively. These diffusivity differences are also represented by the the observed solute supersaturations, $\Delta C_i$, **Fig. 6**. The large initial Al-depletion in the γ(f.c.c.)-matrix reflects the fact that Al is the fastest-diffusing species, which is important for the nucleation and growth of γ′($L1_2$)-precipitates. Rhenium, which has the smallest diffusivity in nickel, is supersaturated in both the γ(f.c.c.)- and γ′($L1_2$)-phases: when it is approaching approximately at one-half its equilibrium concentration the Al concentration is close to its equilibrium value. The different diffusivities of each element for the nucleation and growth kinetics make the compositional trajectories complex in the nucleation and growth stages, leading to compositional variations within the γ′($L1_2$)-precipitates; for example, a concentration gradient within the γ′($L1_2$)-precipitate in a Ni-Al-Cr-Ta alloy [88], hierarchical precipitate structures [89-91], and metastable core-shell nano-precipitates in the Al-Sc system [92-95]. Detailed studies of the nucleation and growth kinetics are needed at significantly shorter initial time intervals: herein, we focus, however, mainly on the coupling effects of the compositional trajectories in the quasi-stationary coarsening regime.

For times greater than 16 h, in the quasi-stationary coarsening regime, the supersaturations, $\Delta C_i$, are very small, $d\Delta C_i / dt \rightarrow 0$, in both phases, and the compositional trajectories are close to being vectors, which agrees with quasi-stationary diffusional behavior [38]. We use the term quasi-stationary because all the concentrations are changing very slowly with time, Table 2. The compositional trajectory as a function of aging time in the γ(f.c.c.)-matrix is fitted using solid-green squares in **Fig. 7,** employing the PV model [38]: $\langle \Delta C_i(t) \rangle = k_i t^{-1/r}$ in Eqn. (6), where the rate constant for the supersaturation of the γ(f.c.c.)-matrix is given by:

$$k_i^{\gamma} = (3\sigma^{\gamma/\gamma'} V_m^{\gamma'})^{2/3} \frac{[(\mathbf{\Delta \bar{C}}^{\gamma\text{-}\gamma'})^T \mathbf{M}_{\gamma}^{-1} \mathbf{\Delta \bar{C}}^{\gamma\text{-}\gamma'}]^{1/3}}{(\mathbf{\Delta \bar{C}}^{\gamma\text{-}\gamma'})^T \mathbf{G}_{\gamma}'' \mathbf{\Delta \bar{C}}^{\gamma\text{-}\gamma'}} \mathbf{\Delta \bar{C}}^{\gamma\text{-}\gamma'} \tag{10}$$

The matrix supersaturations in the PV model are on the tie-line, connecting the equilibrium compositions between the γ(f.c.c.)- and γ′($L1_2$)-phases, because the rate constant for a supersaturation, $k_i^{\gamma}$, is a product of a time-independent scalar and the compositional vector in equilibrium, $\mathbf{\Delta \bar{C}}^{\gamma\text{-}\gamma'}$. The mobility tensor and the Hessian of the Gibbs free energy, $\mathbf{M}_{\gamma}$ and $\mathbf{G}_{\gamma}''$, respectively, affect the magnitude of the supersaturations of the γ(f.c.c.)-matrix from their equilibrium values, but not the direction of the vector. Therefore, the predicted compositions of the γ(f.c.c.)-matrix in the PV model changes along the compositional vector in equilibrium, $\mathbf{\Delta \bar{C}}^{\gamma\text{-}\gamma'}$, which is the direction of the equilibrium tie line. The time-dependent compositions of the γ′($L1_2$)-precipitates do not, however, coincide with the equilibrium tie-line. The rate constants for the supersaturations of the γ′($L1_2$)-precipitates, $k_i^{\gamma'}$, are given by:

$$k_i^{\gamma'} = (3\sigma^{\gamma/\gamma'} V_m^{\gamma'})^{2/3} [(\mathbf{\Delta \bar{C}}^{\gamma\text{-}\gamma'})^T \mathbf{M}^{-1} \mathbf{\Delta \bar{C}}^{\gamma\text{-}\gamma'}]^{1/3} \times \left( \frac{(\mathbf{G}_{\gamma'}''^{-1} \mathbf{G}_{\gamma}'' \mathbf{\Delta \bar{C}}^{\gamma\text{-}\gamma'})}{(\mathbf{\Delta \bar{C}}^{\gamma\text{-}\gamma'})^T \mathbf{G}_{\gamma}'' \mathbf{\Delta \bar{C}}^{\gamma\text{-}\gamma'}} - \frac{\mathbf{G}_{\gamma'}''^{-1} \Delta \bar{\mathbf{V}}}{V_m^{\gamma'}} \right) \tag{11}$$

where $\Delta \bar{\mathbf{V}}$ is the partial molar volume change, $\mathbf{G}_{\gamma'}''$ is the second derivative of the Gibbs free energies with respect to the Ni concentrations in the γ′($L1_2$)-phase. The value of $\Delta \bar{\mathbf{V}}$ is set equal to zero by assuming equal partial molar volumes, and the second derivatives of the Gibbs free energies of the γ′($L1_2$)-phase, $\mathbf{G}_{\gamma'}''$, are obtained using the TCNi8 database employing Thermo-Calc [64]:

$$\mathbf{G}''_{\gamma'} = \begin{pmatrix} \frac{\partial G^2_{\gamma'}}{\partial^2 C_{Al}} & \frac{\partial G^2_{\gamma'}}{\partial C_{Al}\partial C_{Cr}} & \frac{\partial G^2_{\gamma'}}{\partial C_{Al}\partial C_{\mathrm{Re}}} \\ \frac{\partial G^2_{\gamma'}}{\partial C_{Cr}\partial C_{Al}} & \frac{\partial G^2_{\gamma'}}{\partial^2 C_{Cr}} & \frac{\partial G^2_{\gamma'}}{\partial C_{Cr}\partial C_{\mathrm{Re}}} \\ \frac{\partial G^2_{\gamma'}}{\partial C_{\mathrm{Re}}\partial C_{Al}} & \frac{\partial G^2_{\gamma'}}{\partial C_{\mathrm{Re}}\partial C_{Cr}} & \frac{\partial G^2_{\gamma'}}{\partial^2 C_{\mathrm{Re}}} \end{pmatrix} = \begin{pmatrix} 3.05\times10^5 & 1.65\times10^5 & 2.07\times10^5 \\ 1.76\times10^5 & 4.82\times10^5 & 1.93\times10^5 \\ 2.09\times10^5 & 1.85\times10^5 & 10.2\times10^5 \end{pmatrix} \ \mathrm{J\cdot mol^{-1}} \quad (12)$$

From Eqns. 10-12, the calculated rate constants of the supersaturation, $k_i^{PV}$, for Al, Cr, and Re are 0.027 ± 0.006, -0.014 ± 0.003, and -0.004 ± 0.007 mole-fr.s$^{1/r}$ in the γ(f.c.c)-matrix, and 0.026 ± 0.007, -0.007 ± 0.003 and -0.003 ±0.001 mole-fr.s$^{1/r}$ in the γ′($L1_2$)-precipitates, respectively. The calculated $k_i^{PV}$ are approximately 10-20 times smaller than the measured rate constants of the supersaturations, **Fig. 6**, implying that the experimental decays of the supersaturations are significantly larger than the PV model predictions. This can be explained by the faster effective solute diffusivities in the high-volume fraction of γ′($L1_2$)-precipitates.

The compositional trajectories are calculated from the PV model utilizing Eqns. (8-12) and are represented by small green solid-squares, **Fig 7**. In the PV model, the compositional trajectories of the γ(f.c.c.)-matrix lie on the straight dashed tie-line (a vector) connecting the equilibrium compositions of the γ(f.c.c.)- and γ′($L1_2$)-phases at infinite time. The supersaturation of the γ′($L1_2$)-phase is also represented as a vector. It does not, however, coincide with the equilibrium tie-line because the operator $\mathbf{G}''^{-1}_{\gamma'}\mathbf{G}''_{\gamma}$ is applied to $\Delta\bar{\mathbf{C}}^{\gamma\text{-}\gamma'}$ in Eqn. 11, which rotates the direction of the vector.

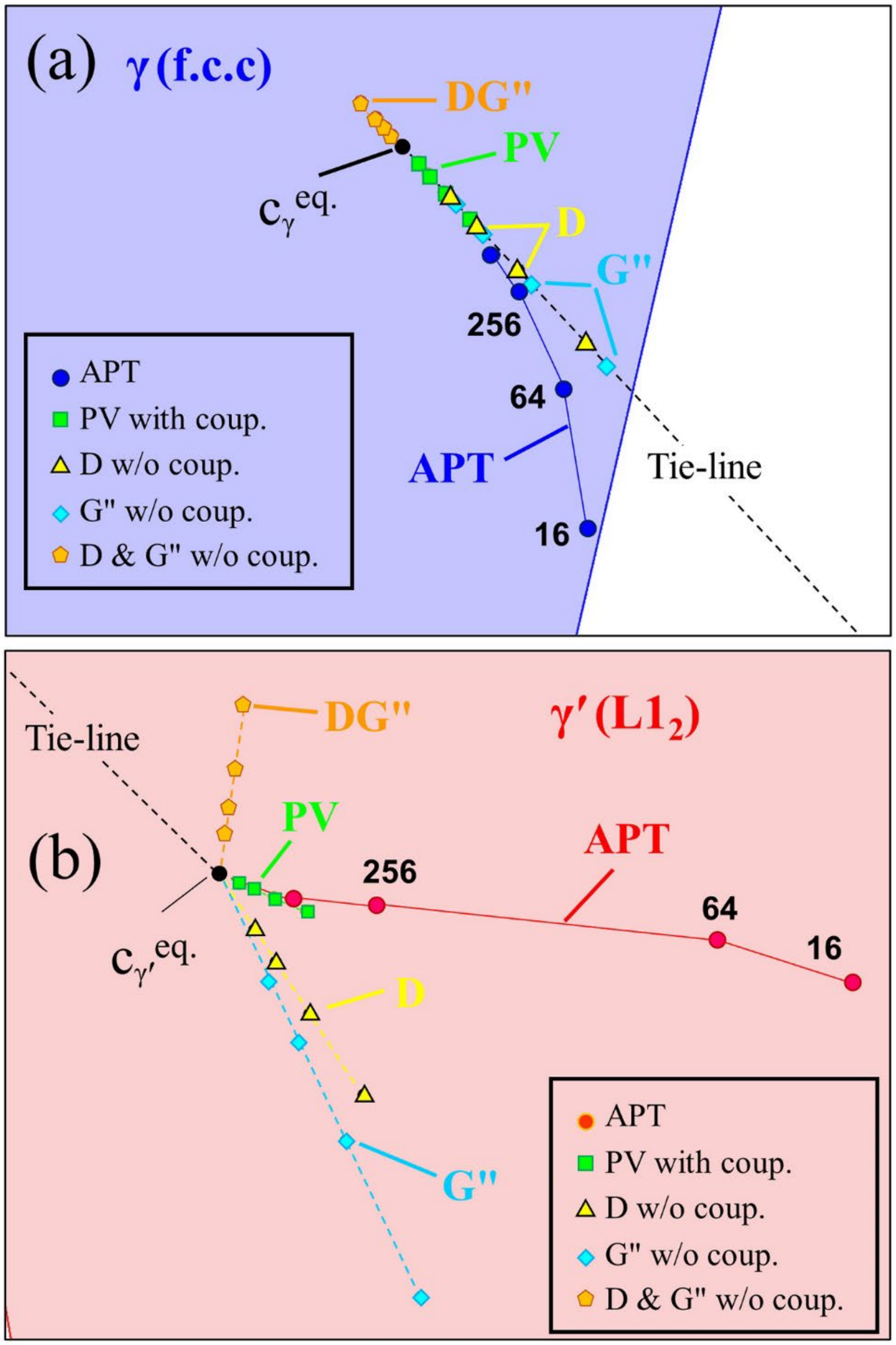

**Fig. 8.** A graphical representation of the compositional trajectories in the coarsening regime (t ≥ 16 h) with and without the coupling terms of the $\mathbf{G}''$ and $\mathbf{D}$ tensors. The equilibrium tie-line, $\Delta\bar{C}^{\gamma\text{-}\gamma'}$, is indicated by a black dashed-line in (a) and (b). The APT experimental data points are represented by the solid-blue- and solid-red-circles for: (a) the γ(f.c.c.)- and (b) γ′($L1_2$)-phases, respectively. The compositional trajectories calculated from the PV model commencing at an aging time of 16 h are represented by solid-green-squares (*all terms in* $\mathbf{G}''$ *and* $\mathbf{D}$ *are included*), yellow colored-triangles (*without the off-diagonal terms in* $\mathbf{D}$), cyan colored-diamonds (*without*

*the off-diagonal terms in* $\mathbf{G}''$), and orange-colored pentagons (*without the off-diagonal terms in* $\mathbf{G}''$ *and* $\mathbf{D}$).

The diffusional coupling effects on the directions of the compositional trajectories in the coarsening regimes (t ≥ 16 h) are represented in **Fig. 8**. The detailed compositional trajectories with or without the off-diagonal terms of the $\mathbf{G}''$ and $\mathbf{D}$ tensors are represented in magnified figures, **Fig. 8(a)** and **Fig. 8(b)** for the γ(f.c.c.)- and γ′($L1_2$)-phases, respectively. The light-blue- and light-red-colored surfaces represent the calculated conjugate solvus surfaces for the γ(f.c.c.)- and γ′($L1_2$)-phases, respectively. The equilibrium compositions of the γ(f.c.c.)- and γ′($L1_2$)-phases at infinite time are indicated by two solid-black circles in **Figs. 8(a,b)**, and the experimental data points (APT) in the coarsening regime (t ≥ 16 h) are indicated by solid-blue- and solid-red-circles for the γ(f.c.c.)- and γ′($L1_2$)-phases, respectively. The compositional trajectories calculated from the PV model, commencing at an aging time >16 h are represented by solid-green squares (*with all the terms in* $\mathbf{G}''$ *and* $\mathbf{D}$ *included*), yellow-triangles (*without including the off-diagonal terms in* $\mathbf{D}$), cyan-diamonds (*without including the off-diagonal terms in* $\mathbf{G}''$), and orange-pentagons (*without including the off-diagonal terms in both* $\mathbf{G}''$ and $\mathbf{D}$). The compositional trajectories of the γ(f.c.c.)-matrix in the PV model *with and without the coupling terms of* $\mathbf{G}''$ *and* $\mathbf{D}$ *tensors*, **Fig. 8(a)**, lie on a straight black dashed tie-line connecting the equilibrium compositions of the γ(f.c.c.)- and γ′($L1_2$)-phases, $C_i^{\gamma,eq}$ and $C_i^{\gamma',eq}$, respectively.

The magnitudes of the solute supersaturations, $\Delta C_i$, also change with or without the off-diagonal terms in both $\mathbf{G}''$ and **D.** The compositional trajectories of the γ(f.c.c.)-matrix without the off-diagonal terms in $\mathbf{G}''$ [cyan-colored-diamonds in **Fig. 8(a)**] has the largest value of $\Delta C_i$ and is compatible with the APT results (solid blue-circles) after 256 h. The calculated supersaturations without the off-diagonal terms in both the $\mathbf{G}''$ and $\mathbf{D}$ tensors [orange-colored-pentagons in **Fig. 8(a)**] are in opposite directions along the tie-line around the origin, which is related to the negative value of the interfacial energy, -18.5 mJ/m$^2$. A negative value of the interfacial free energy is, however, impossible physically as discussed above.

The supersaturations of the γ′($L1_2$)-phase in the PV model is also a vector in **Fig 8(b)**, solid-green colored squares, albeit the direction deviates from the equilibrium tie-line due to application of the operator $\mathbf{G}_{\gamma'}^{''-1}\mathbf{G}_{\gamma}^{''}$ to the compositional vector in equilibrium, $\Delta\bar{\mathbf{C}}^{\gamma-\gamma'}$. The direction of the compositional trajectories of the γ′($L1_2$)-phase in the PV model without the off-diagonal terms in both **G″** and **D** (*orange-colored pentagons in* **Fig. 8b**) are far from the equilibrium tie-line vector as well as the direction of the APT experimental data. The directions without the off-diagonal terms in **G″** [*yellow-colored triangles in* **Fig. 8(b)**] and in **D** [*cyan colored-diamonds in* **Fig. 9(b)**] are close to the equilibrium tie-line vector: they are, however, far away from the direction of the APT data. As indicated by the solid-green squares, the direction of the compositional trajectories with all off-diagonal terms in both the **G″** and **D** tensors are close to the APT experimental data, **Fig. 8(b)**, confirming the importance of the off-diagonal terms in this case.

The compositional paths of the γ(f.c.c.)- and γ′($L1_2$)-phases for the APT results and PV coarsening model are successfully captured and compared. Including the coupling (off-diagonal terms) in the Hessian of the Gibbs free energies, **G″**, and the diffusion tensor, **D**. The interfacial free energies and the compositional trajectories in the quasi-stationary coarsening regime are correctly predicted and in agreement with our APT data. The **G″** and **D** values do not include, however, the role of vacancies on diffusion phenomena because vacancies are assumed to be in thermodynamic equilibrium in Thermo-Calc and DICTRA: that is, the chemical potential of the vacancies are identically equal to zero [37, 76]. Vacancy-mediated lattice-kinetic Monte Carlo (LKMC) simulations with parameters deduced from first-principles calculations can account correctly for vacancy diffusion mechanism in nickel-based binary Ni-Al [36] and ternary Ni-Al-Cr alloys [37]. The role of vacancy-solute binding energies and their effects on flux coupling utilizing vacancy-mediated LKMC simulations and first-principles calculations are suggested for future studies of the present quaternary alloy.

## 5. Summary and conclusions

The effects of diffusional coupling on the compositional trajectories and the interfacial Gibbs free energies of γ(f.c.c.)/γ’($L1_2$)-heterophase interfaces during phase separation (i.e., precipitation or phase decomposition) in a quaternary Ni–0.1Al-0.85Cr-0.02Re (mole-fraction) alloy aged at 700 °C from 0 h to 1024 h, were studied by atom-probe tomography (APT), transmission electron microscopy (TEM), and the Philippe-Voorhees (PV) phase-field coarsening model. The temporal evolution of the microstructures and the compositional trajectories are measured employing APT and TEM experiments, and the roles of the off-diagonal diffusion terms in the $\mathbf{G}''$ (the second derivatives of the molar Gibbs free energies) and $\mathbf{D}$ (diffusivity) tensors with respect to the PV phase-field model are compared to the experimental data, leading to the following results and conclusions:

1. The mean precipitate radius, $\langle R(t) \rangle$, number density, $N_v(t)$, and precipitate volume fraction, $\phi_{\gamma'}$, during temporal evolution of γ′(L1₂)-precipitates are measured by TEM and APT, **Figs. 1** and **2**. The peak number density of the γ′(L1₂)-precipitates, $(8.3 \pm 0.8)\cdot 10^{23}$ m$^{-3}$, occurs at 0.25 h, and this first-order phase transformation enters a concomitant growth and coarsening regime between 4 h and 16 h. The precipitate volume fraction, $\phi_{\gamma'}(t)$, measured by APT is close to zero at 0 h, [$\boldsymbol{\phi_{\gamma'}(t = 0\ h)}$ = (0.38 ± 0.05) %], which is asymptotically approaching a constant value after 16 h, [$\phi_{\gamma'}(t = 16h)$ = (36.71 ± 3.6) %]. During the quasi-stationary coarsening regime, evaluated at t ≥ 16 h, the temporal power-law exponent for $\langle R(t) \rangle$ is 0.32 ± 0.03, and for $N_v(t)$ it is -0.87 ± 0.12, **Fig. 3**.
2. The APT results demonstrate that a high fraction of the γ′(L1₂)-precipitates is interconnected by necks, $f(t)$ = (45.83 ± 6.12) %, with small mean edge-to-edge distances between γ′(L1₂)-precipitates, $\langle \lambda(t) \rangle$ = (9.19 ± 4.32) nm, after aging for 1 h, **Fig. 4**. Large fractional values of $f(t)$ and small values of $\langle \lambda(t) \rangle$ are consistent with the coagulation-coalescence coarsening mechanism as opposed to the evaporation-condensation ("the big eat the small") mechanism, which is implicit in all coarsening models starting with the Lifshitz-Slyozov (LS) diffusion-controlled model, without experimental proof. For aging times greater than 16 h, the evaporation-condensation mechanism commences to dominate over the coagulation-coalescence mechanism, as $\langle \lambda(t) \rangle$ is > 16 nm when $f(t)$ is < 20 %. At 1024 h $\langle \lambda(t) \rangle$ is 32.84 ± 12.03 nm and $f(t)$ is 3.57 ± 1.015, which implies that the evaporation-condensation mechanism is dominating.

3. The compositions of the γ(f.c.c.)-matrix and the γ′($L1_2$)-precipitates evolve temporally, **Figs. 5 and 6**. The composition of the γ(f.c.c.)-matrix after vacancy annealing and quenching is a solute-supersaturated composition of 0.7819Ni-0.1051Al-0.0921Cr-0.021Re mole-fr., while the γ′($L1_2$)-nuclei have the composition 0.768Ni-0.1944Al-0.0625Cr-0.0163Re mole-fr. at $\langle R(t_o)\rangle$ = (1.06 ± 0.25) nm. As the γ′($L1_2$)-precipitates grow, excesses of Ni, Cr and Re, and the depletion of Al in the γ(f.c.c.)-matrix develops because of the diffusional fluxes crossing the γ(f.c.c.)/γ′($L1_2$)-heterophase interfaces. The equilibrium γ(f.c.c.)-matrix and γ′($L1_2$)-precipitate compositions at *infinite time* are estimated to be 0.8015Ni-0.0524Al-0.1173Cr-0.0288Re mole-fr. for the γ(f.c.c.)-matrix, and 0.7588Ni-0.1762Al-0.0543Cr-0.0107e mole-fr. for the γ′($L1_2$)-precipitates.

4. The coarsening kinetics of the γ′($L1_2$)-precipitates, mean precipitate radius, $\langle R(t)\rangle$, number density, $N_v(t)$, and solute supersaturations ($\Delta C_i$), are compared with a diffusion-limited phase-field coarsening model developed by Philippe and Voorhees (PV). The mobility tensor, **M**, in the PV model, in combination with a thermodynamic factor, yields diffusivities through the relationship, $\mathbf{D} = \mathbf{M}\mathbf{G}''$, where $\mathbf{G}''$ is the second derivative (Hessian) of the Gibbs free molar energies of the γ(f.c.c.)-phase. Using Ni as the reference component, the $\mathbf{G}''$ and $\mathbf{D}$ tensors for both phases were computed utilizing the thermodynamic database TCNi8 and the Ni-mobility database from Thermo-Calc at their equilibrium compositions at 700 °C [41, 64].

5. The calculated value of the Gibbs interfacial free energy, $\sigma^{\gamma/\gamma'}$, is strongly influenced by the off-diagonal terms in the $\mathbf{D}$ and $\mathbf{G}''$ tensors, which are included in the PV model. The value of $\sigma^{\gamma/\gamma'}$, based on the APT experiments and the PV model calculations at 700 °C are: (i) (6.9 ± 1.41) mJ/$m^2$ including the off-diagonal terms in the $\mathbf{D}$ and $\mathbf{G}''$ tensors; (ii) (18.9 ± 2.1) mJ/$m^2$ *without the off-diagonal terms* in the $\mathbf{D}$ tensor; (iii) (37.7 ± 3.3) mJ/$m^2$ *without the off-diagonal terms* in the $\mathbf{G}''$ tensor; and (iv) (-7.5 ± 1.2) mJ/$m^2$ *without the off-diagonal terms* in the $\mathbf{D}$ and $\mathbf{G}''$ tensors. A negative value of $\sigma^{\gamma/\gamma'}$ is nonphysical because the system will continuously generate interfaces until classical thermodynamics is no longer applicable as noted by J. W. Gibbs in 1876 [83].

6. The compositional trajectories of the γ(f.c.c.)-matrix and γ′($L1_2$)-precipitate phases are represented on a partial Ni-Al-Cr-Re quaternary phase diagram employing a partial tetrahedron,

**Fig. 7.** In the nucleation and growth regimes of the nanoprecipitates, t ≤ 4 h, the small diffusivities of Cr and Re compared to Al's large diffusivity, results in a curvilinear trajectory, which deviates from the equilibrium γ(f.c.c.)-γ′($L1_2$) tie-line between the γ(f.c.c.)-matrix- and the γ′($L1_2$)-precipitate phases. In the quasi-stationary coarsening regime, t ≥16 h, the supersaturations, $\Delta C_i$, are small in both phases, and the compositional trajectories are close to being a vector, which agrees with the quasi-stationary diffusional behaviors described by the diffusion-limited PV model for both the γ(f.c.c.)-matrix and γ′($L1_2$)-precipitate-phases.

7. The compositional trajectories measured from our APT experiments are compared with the PV coarsening model in the quasi-stationary coarsening regime (t ≥ 16 h) with and without the off-diagonal terms of the $\mathbf{G}''$ and $\mathbf{D}$ tensors, **Fig. 8.** The calculated compositional trajectories of the γ(f.c.c.)-matrix in the PV model (pure coarsening) with and without the coupling terms of the $\mathbf{G}''$ and $\mathbf{D}$ tensors lie on a vector connecting the equilibrium compositions of the γ(f.c.c.)- and γ′($L1_2$)-phases on the two calculated conjugate solvus surfaces. The compositional trajectory of the γ′($L1_2$)-phase in the PV model is a vector; its direction, however, does not coincide with the equilibrium tie-line. The direction of the compositional trajectory of the γ′($L1_2$)-phase in the PV model with the off-diagonal terms in both the $\mathbf{G}''$ and $\mathbf{D}$ tensors agrees with the direction of the compositional trajectory measured by our APT experiments after 16 h of aging.

8. In the Supplemental Materials Sections, we compare nonlinear multivariate regression (NLMR) analyses with linear multivariate regression (LMR) analyses to obtain accurate values of the temporal exponents for the mean radius, *p*, and the temporal exponent for the supersaturations, *r*. The optimized value of *p* using the LMR analyses for the current APT and TEM data is smaller than in the NLMR analyses (2.64 vs. 3.06) due to the strong dependency on the largest value with the temporal factor of *p* for linearization. It is the same for the evaluation of the exponents for the supersaturations, *r*. The smaller exponent value in the LMR analyses than in the diffusion-limited LS and PV models was used to claim the validity of the trans-interface diffusion controlled (TIDC) model [96-98]; it is, however, a misleading approach for analyzing experimental coarsening data and is also a poor predictor of the coarsening mechanism.

9. The TDIC model also posits that the interfacial width, δ(t), increases with increasing <R(t)>. Our APT measurements are inconsistent with this ansatz of the TDIC model. That is, we observe experimentally that with an increasing mean radius, $\langle R(t) \rangle$, the normalized interfacial width, $\frac{\delta(t)}{\langle R(t) \rangle}$, decreases with increasing aging time through 1024 h: this ratio is the relevant physical quantity. And this ratio is proportional, to first order, to $\langle R(t) \rangle^{-1}$. The decrease of δ(t) with increasing <R(t)> is also consistent with decreases in the interfacial widths in binary Ni-Al alloy systems [99], ternary [100], and multi-component alloys [101, 102].

## Acknowledgments

This research was supported by the National Science Foundation, Division of Materials Research (DMR) grant number DMR-1610367 001, Profs. Diana Farkas and Gary Shiflet, grant officers. Atom-probe tomography was performed at the Northwestern University Center for Atom-Probe Tomography (NUCAPT), which is a core facility of Northwestern University. The LEAP500XS tomograph at NUCAPT was purchased and upgraded with grants from the NSF-MRI (DMR-0420532) and ONR-DURIP (N00014-0400798, N00014-0610539, N00014-0910781, N00014-1712870) programs. This work made also use of the MatCI Facility and EPIC facility of Northwestern University's NUANCE Center. NUCAPT, MatCI, and EPIC facilities receive support from the MRSEC program (NSF DMR-1720139), SHyNE Resource (NSF ECCS-2025633), the IIN, and the Initiative for Sustainability and Energy (ISEN) at Northwestern University. The synchrotron x-ray diffraction measurements were performed at the Advanced Photon Source at Argonne National Laboratory, Lemont, IL. We thank research professor Dennis Keane for his important help with these measurements. SB and ZM thank Dr. Fan Zhang, Dr. Jun Zhu, Dr. Duchao Lv, and Dr. Weisheng Cao at Computherm LLC (Madison, WI) for their help in representing the quaternary phase diagram. DNS thanks Georges Martin (CEA Saclay, France) and Peter Voorhees (Northwestern) for many illuminating discussions concerning coarsening (Ostwald ripening) and mean-field modeling in general during many years. We also thank research associate professor Dr. Dieter Isheim for managing NUCAPT.

**Supplementary Materials**

# The effects of diffusional couplings on compositional trajectories and interfacial free energies during phase separation in a quaternary Ni-Al-Cr-Re model superalloy

Sung-Il Baik[1,2], Zugang Mao[1], Qingqiang Ren[1], Fei Xue[1], Carelyn E. Campbell[3], Chuan Zhang[4], Bicheng Zhou[1,5], Ronald D. Noebe[6], David N. Seidman[1,2*]

[1.] Department of Materials Science and Engineering, Northwestern University, Evanston, IL 60208, USA.
[2.] Northwestern University Center for Atom Probe Tomography (NUCAPT), Evanston, IL 60208, USA
[3.] Materials Science and Engineering Division, National Institute of Standards and Technology (NIST), 100 Bureau Dr. Gaithersburg, MD 20899-8555, USA
[4.] Computherm LLC, 8401 Greenway Blvd. Suite 248, Middleton, WI 53562, USA
[5.] Department of Materials Science and Engineering, University of Virginia, Charlottesville, VA, 22904, USA
[6.] NASA Glenn Research Center, 21000 Brookpark Rd, Cleveland, OH 44135, USA

**Corresponding authors email addresses: si-baik@northwestern.edu (Sung-Il Baik)*
*d-seidman@northwestern.edu* (David N. Seidman)

## Supplement A: Full 3D APT reconstructions for the microstructural evolution of γ′($L1_2$)-precipitates

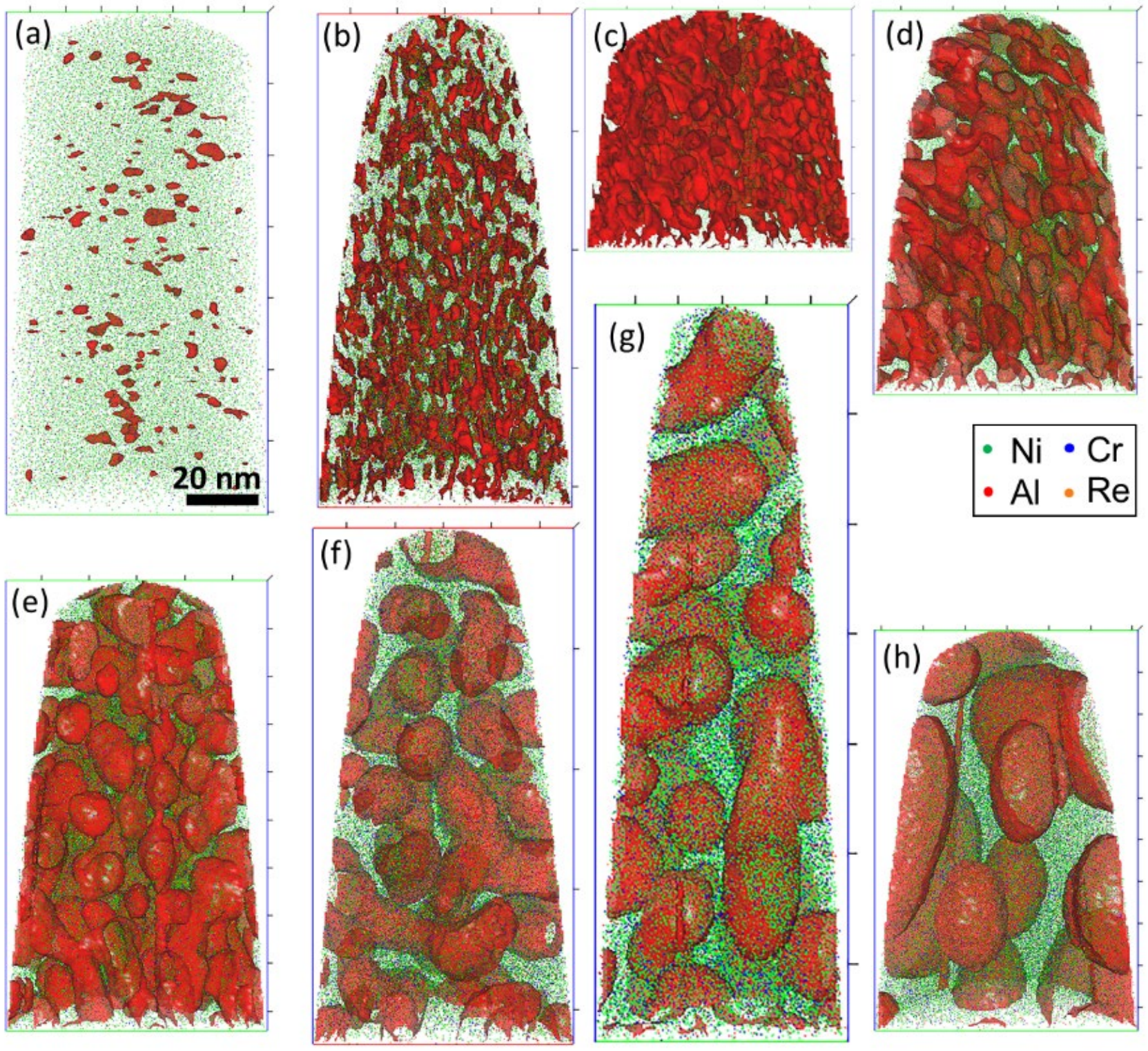

**Fig. S1.** 3-D APT reconstructions of the γ(f.c.c.)-matrix and γ′($L1_2$)-precipitates in the quaternary Ni-0.10Al-0.085Cr-0.02Re alloy aged for: (a) 0 h; (b) 0.25 h; (c) 1 h; (d) 4; (e) 16 h; (f) 64 h; (g) 256 h; and (h) 1024 h at 973 K (700 ºC). Only a fraction (0.2%) of the Ni (green), Al (red), Cr

(blue), and Re (dark yellow) atoms are displayed, for the sake of clarity, and the γ(f.c.c.)/γ′($L1_2$)-heterophase interfaces are delineated by red 0.14 mole-fr. Al iso-concentration surfaces. The lozenge-shaped precipitates are further evidence for the coagulation-coalescence mechanism of coarsening (Ostwald ripening) as opposed to the evaporation-condensation coarsening ("the large eat the small") mechanism, which is implicit in the LS model for coarsening (Ostwald ripening). Note how the γ′($L1_2$)-precipitates become more lozenge-shaped with increasing aging time, (f), (g), (h).

**Supplement B: Synchrotron X-ray diffraction (XRD) analysis for determining the lattice parameter misfit.**

**Fig. S2** displays high-temperature synchrotron x-ray diffraction (Advanced Photon Source at Argonne National Laboratory) measurements of the Ni-0.10Al-0.085Cr-0.02Re alloy at 700 °C for an aging time of 1024 h. Lattice parameter measurements at ambient and annealing temperatures were performed using synchrotron X-ray diffraction (XRD) (λ = 0.620422 Å) at the Advanced Photon Source (APS) at Argonne National Laboratory, Lemont, Illinois. A beam size of (3.5 × 0.5) $mm^2$ was employed, and at least one family of peaks was detected, i.e., (003) and (004), for the γ(f.c.c.)-γ′($L1_2$) peak deconvolution procedure and for lattice parameter misfit calculations employing the Bragg-Brentano geometry with a moving sample-stage and point detector at room temperature and 700 °C. Samples aged at 700 °C for 1024 h were mounted in an Anton Paar DHS 1100 hot stage with a graphite dome to minimize X-ray absorption in a vacuum of < $3.8 \times 10^{-2}$ Pa prior to heating. A heating rate of 100 $°C \cdot min^{-1}$ was utilized to increase the temperature to 600 °C, and 50 $°C \cdot min^{-1}$ to 700 °C, and then held for at least 20 min before scanning. A pseudo-Voigt function [1] was employed to fit the detected peaks with two peaks for the γ(f.c.c.)-and γ′($L1_2$)-phases employing the computer program *Origin*.

The lattice parameter of the γ′($L1_2$)-precipitate, $a_{\gamma'}$, is 0.3587 nm at 700 °C, based on the (003) superlattice reflection of the γ′($L1_2$)-phase, **Fig. S2(a)**. Whereas the lattice parameter of the γ(f.c.c.)-matrix, $a_\gamma$, is determined by employing a deconvolution of the (004) diffraction peaks of the γ′($L1_2$)- and γ(f.c.c.)-phases. The deconvolution fitting procedure of the data, which is represented as a red curve in **Fig S2(b)**, is well characterized by the green curve for the γ(f.c.c.)-phase and the blue curve for the γ′($L1_2$)-phase. The lattice parameter of the γ(f.c.c.)-matrix, $a_\gamma$, at 700 °C is 0.3591 nm, and the lattice parameter misfit, δ, is then calculated from the following equation:

$$\delta = \left( \frac{2(a_{\gamma'} - a_\gamma)}{a_{\gamma'} + a_\gamma} \right) \quad \text{(S1)}$$

The extracted values from the synchrotron XRD measurements at 700 °C are 0.3591 nm and 0.3587 nm for the γ(f.c.c.) and γ′($L1_2$) phases, respectively, resulting in a lattice-parameter misfit, δ, of -0.11%. This negative mismatch means that the γ′($L1_2$)-precipitate has a smaller lattice parameter than the matrix, which is due to the large Re atoms partitioning to the γ(f.c.c.)-matrix [2]. Vegard's rule coefficients for Re are reported to be 0.441 and 0.262 for the γ(f.c.c.)- and γ′($L1_2$)-phases [3], respectively. The large partitioning coefficient of Re to the γ(f.c.c.) matrix (~ 2.63) results in a negative value for the lattice-parameter misfit. The lattice parameter misfit, δ, has positive values for Ni-A1 and Ni-Al-Cr alloys: + 0.57 % for the Ni-A1 binary system [4] and 0.06 ± 0.07, and 0.22 ± 0.07% for Ni-0.052 Al-0.142 Cr mole-fr., and for Ni-0.098 Al-0.083 Cr mole-fr. ternary alloys, respectively [5, 6]. The elastic energy is proportional to the square of the lattice parameter misfit, δ, and its magnitude governs the elastic energy interactions among the γ′($L1_2$) precipitates, which potentially influences the entire coarsening behavior. The small negative value of δ obtained by adding Re to Ni-based superalloys to retard the coarsening kinetics of γ′($L1_2$)-precipitate, which yields longer creep-rupture lifetimes [7].

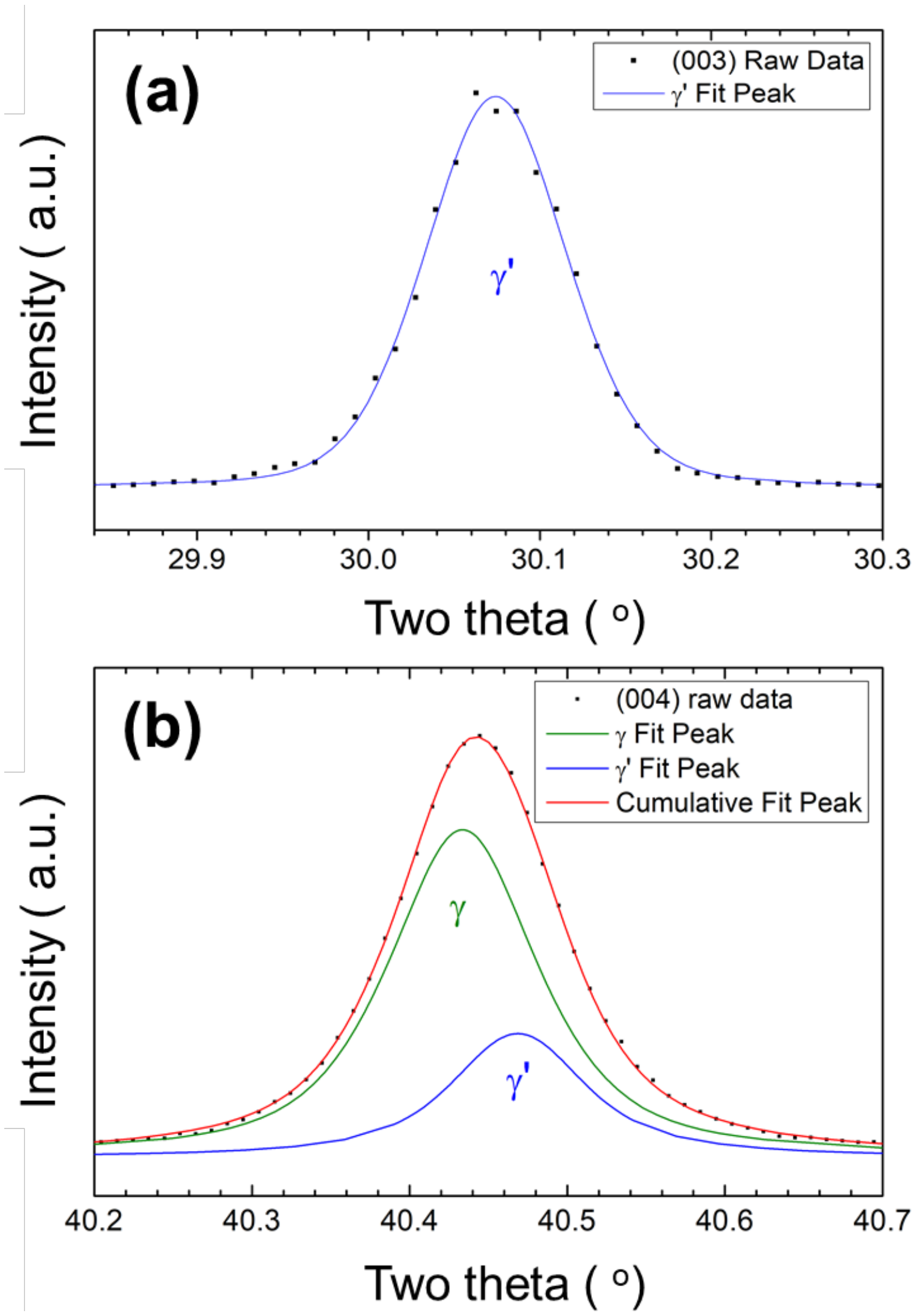

**Fig. S2.** High-temperature synchrotron XRD patterns at 700 °C of: (a) (003) peak; and (b) (004) peak of the Ni-0.85Cr-0.1Al-0.02Re alloy for an aging time of 1024 h. The lattice parameter of the γ′($L1_2$)-phase, $a_{\gamma'}$, is measured from the (003) superlattice reflection, and the lattice parameter of the γ(f.c.c.)-matrix, $a_\gamma$, is determined by deconvoluting the cumulative (004) diffraction peaks of the γ′($L1_2$)- and γ(f.c.c.)-phases: the blue curve is for the γ′($L1_2$)-phase and the green curve is for γ(f.c.c.)-matrix. The lattice parameters of the γ(f.c.c.)- and γ′($L1_2$)-phases, $a_\gamma$ and $a_{\gamma'}$, at 700 °C are 0.3591 nm and 0.3587 nm, respectively, resulting in a lattice-parameter misfit, δ, of -0.11%.

## Supplement B: Calculating Temporal Exponents from Experimental APT Data

### *B1. Comparison of linear and nonlinear analyses of the temporal exponents of the mean radius, $\langle R(t)\rangle$, for the quaternary Ni–0.10Al-0.085Cr-0.02Re alloy aged at 700 °C (973 K)*

We calculated the temporal exponent, *p*, of the mean radius, <R(t)>, employing Eq. (5), $\langle R(t)\rangle^p - \langle R(t_o)\rangle^p = K(t-t_o)$, using the *nonlinear multivariate regression* (NLMR) methodology [8], and plotted <R(t)> vs. aging time on *linear scales, **Fig. S3***. We emphasize strongly that this basic equation is nonlinear. This plot is presented using a log-log format in the main text, **Fig. 2**, which we don't use, however, to calculate the temporal exponent, $p$. Again, $K$ is the rate constant for coarsening of the mean radius, $\langle R(t)\rangle$, $t_o$ is the time for the onset of stationary or at least quasi-stationary coarsening, and $\langle R(t_o)\rangle$ is the mean radius at a time $t_o$. The temporal exponent, $p$, is calculated using a NLMR analysis of the basic relationship for $\langle R(t)\rangle$, *with no assumptions being made about the four variables:* $p$, *K,* $t_o$, *and* $\langle R(t_o)\rangle$. Although $p$ and $K$ are the most important parameters in Eq. (5), the $t_o$ and $\langle R(t_o)\rangle$ values are significant to determine the correct values of $p$ and $K$. Moreover, these parameters tell us when the coarsening regime model works and what is the <R(t)> value. The SciPy library in the Python module [9] is utilized to solve a nonlinear least-squares problem with bounds on the variables. The goodness of fit of the curved feature is represented by Chi-squared ($\chi^2$) value, which is defined by:

$$\chi^2 = \sum_i \frac{(y_i - f_i)^2}{\sigma_i^2} \qquad \text{(S2)}$$

where $y_i$ is the measured quantity, $\langle R(t)\rangle$ in this case, at a given aging time, and $f_i$ is the associated modeled value from a nonlinear fit: the $\sigma_i^2$ are the squared standard deviations for each data point. The reason for dividing by the squared standard deviations is to reduce the influence of large residuals, which contribute to the robustness of the solution. Sometimes it is also weighted by model values, $f_i$: t-test, and by the degrees of freedom, the number of observed variables minus the number of fitted parameters: i.e., reduced Chi-squared ($\chi^2_{red}$). Herein, only the data for t ≥ 4 h are included in the fitting analysis because the volume fraction of γ′($L1_2$) is asymptotically approaching its equilibrium value. A NLMR analysis yields K = 6.20 ± 2.4) x$10^{-30}$ $m^3s^{-1}$ and a temporal exponent for <R(t)> of p = 0.32 ± 0.03; This value agrees, within error, with the value p = 3, predicted by the diffusion-limited Lifshitz-Slyozov (LS) model for binary alloys and the Philippe-Voorhees (PV) model for multi-component non-dilute alloys. The quantity $t_o$ is 10,525 seconds (2.92 h) and $\langle R(t_o)\rangle$ is 4.23 nm, which are in reasonable in agreement when the coarsening regime commences.

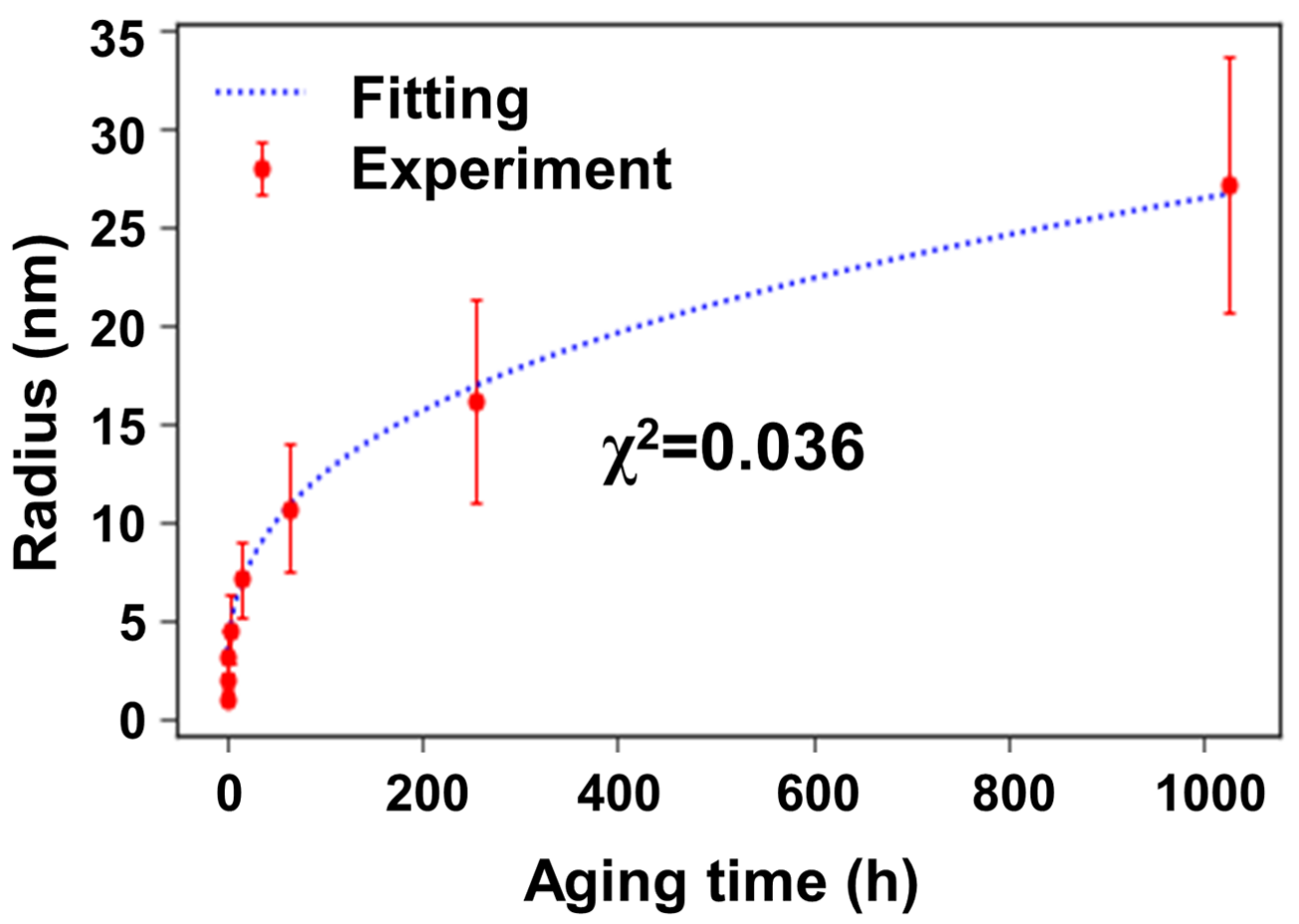

**Fig. S3.** Nonlinear multivariate regression (NLMR) analysis of the temporal exponent, *p*, in $\langle R(t)\rangle^p$ versus time, t, in hours plot. The data is plotted for all the aging times utilized; only the data for t ≥ 4 h are included in the analysis because after 4 h the volume fraction of γ′($L1_2$) is asymptotically approaching its equilibrium value. The goodness of fit is represented by the minimization of Chi-squared ($\chi^2$).

In contrast, the linear plots are derived by plotting $\langle R(t)\rangle^p$ versus time (h) for *different assumed temporal exponents*, *p = 2, 2.6, 3, and 4*, ***Fig. S4***. The goodness of fit to a linear model is obtained from the coefficients of determination, $x^2$, which is based on linearizing Eq. 5. It is invalid, however, for nonlinear models and most statistical software programs can't calculate it. Linear plots were used by Ardell [10, 11] to obtain optimized values of the rate constant, *K*, and the temporal exponent, *p*. We also calculated them in the same fashion for comparative purposes. A coefficient of determination for a linear fit is commonly denoted as an "R_squared ($R^2$) value," and are represented herein as $\xi^2$ to distinguish it from the mean radius, $\langle R(t)\rangle$. The quantity $\xi^2$ is calculated using two different sums of squares, the total sum of squares, $SS_{tot}$, and the sum of square residuals, $SS_{res}$:

$$SS_{tot} = \sum_i \left(y_i - \langle y\rangle\right)^2 \tag{S3}$$

$$SS_{res} = \sum_i \left(f_i - y_i\right)^2 \tag{S4}$$

$$\xi^2 = 1 - \frac{SS_{res}}{SS_{tot}} \tag{S5}$$

where $\langle y\rangle$ is the mean of all measured $y_i$-values, and $\xi^2$ is a number between 0 and 1, which measures how much better is a given model than a control model. The better the linear regression fits the data, the closer the value of $\xi^2$ is to 1. The linear plots of $\langle R(t)\rangle^p$ versus time in second (s) plots for $p$ values of 2, 2.6, 3, and 4, have the largest values of $\xi^2$ with 0.9996 at $p$ = 2.6. A large value of the quantity $\xi^2$ *is not indicative of a good fit in curve fitting* because linear plots of $\langle R(t)\rangle^p$ versus time (h) have a strong dependence on the largest $\langle R(t)\rangle$ due to the power factor of $p$. For example, the largest value of $\xi^2$ with 0.9996 is obtained for $p$ = 2.6, the deviations from a linear fit at 4 and 64 h are clearly visible in the inset graph, **Fig. S4(b)**. This trend is also found in Fig. **S4(a)** and **(d)** for linear fitting at p = 2 and 4; it is, however, minimized at p = 3. The linear fits of $\langle R(t)\rangle^p$ versus time (h) for $p$ = 3 produce, however, negative values when the *time equals zero*; for example, the linear equations for p = 3 and 4, **Fig. S4(c)** and **(d)**, are $\langle R(t)\rangle^3 = 0.0054t - 2.528$ *and* $\langle R(t)\rangle^4 = 0.1408t - 2028.19$, respectively, implying that the y-intercepts of the linear fits have negative values at time zero, which is nonphysical. These deviations at the initial time are underestimated through the power of $p$ for the $\langle R(t)\rangle$ values in the linear fitting model.

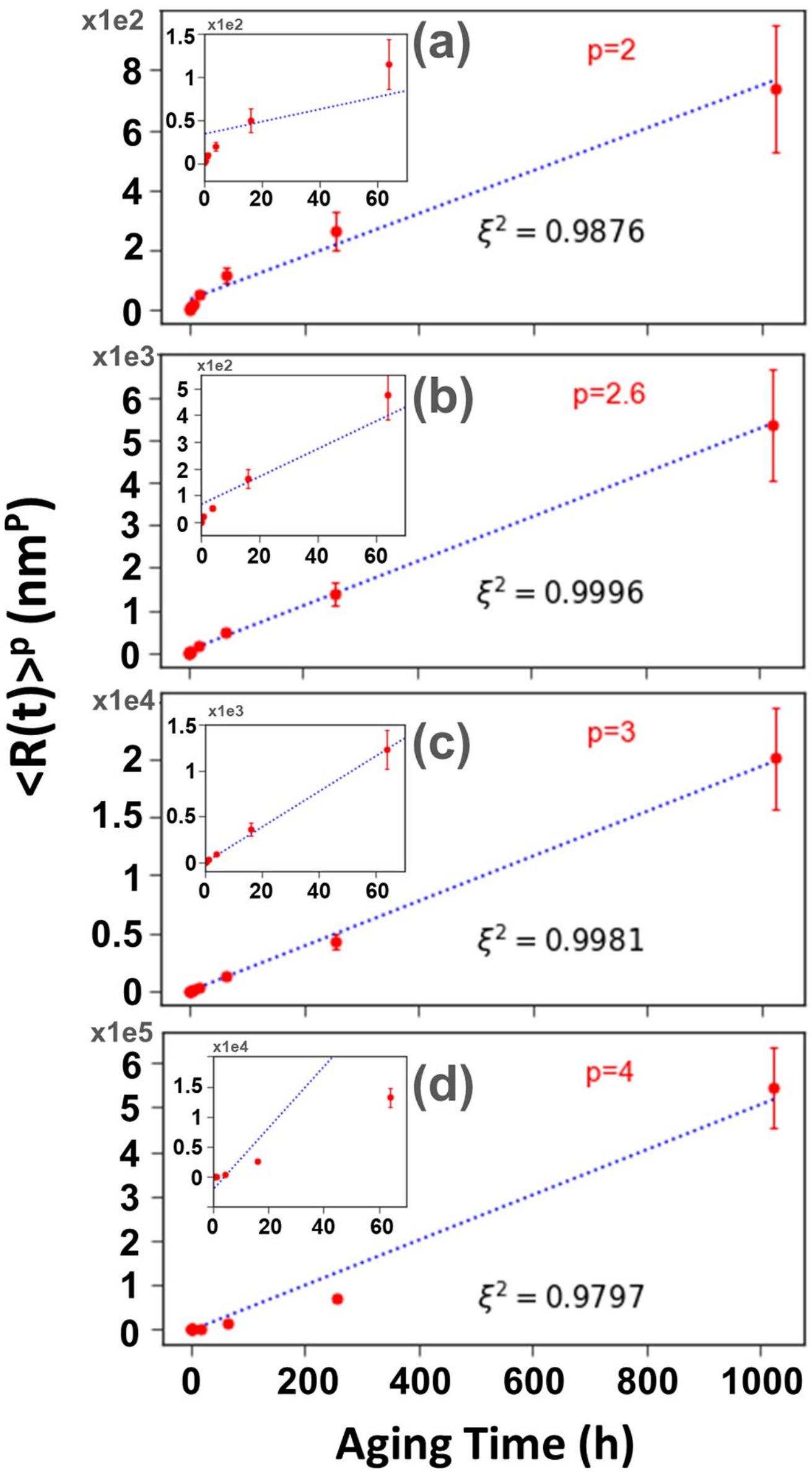

**Fig. S4.** $\langle R(t)\rangle^p$ versus time plots for $p$ values of: (a) $p$ = 2; (b) $p$ = 2.6; (c) $p$ = 3; and (d) $p$ = 4, for their associated linear fits. The coefficient of determination, $\xi^2$, is given for each linear fit to the data.

The other issue for the linear fitting model is that the optimized exponent $p$-values are much smaller than the nonlinear fitting value. **Fig. S5** displays the chi-squared, $\chi^2$, and the coefficients of determination, $\xi^2$, as a function of the temporal exponent $p$. Due to the characteristics of equations S2-S5, Chi-squared ($\chi^2$) employs the minimum value, while the coefficients of determination, $\xi^2$ utilizes the maximum value close to 1. The minimum and maximum values of $\chi^2$ and $\xi^2$ are p = 3.06 and 2.64, respectively. These values are obtained when all the variables, $p$, $K$, $t_o$ and $\langle R(t_o)\rangle$ are included. Without the $t_o$ and $\langle R(t_o)\rangle$ values for fitting (red curves), which Ardell performed [10, 11], the minimum and maximum values of $\chi^2$ and $\xi^2$ are at p = 3.13 and 2.78, respectively, which are slightly larger than the values of fitting with all variables. The $\chi^2$- and $\xi^2$-curves with respect to $p$ have smaller curvatures without the $t_o$ and $\langle R(t_o)\rangle$ because the remaining constant for the linear fitting of $\langle R(t)\rangle^p$ is the rate constant $K$, which limits the goodness of fitting. In the coarsening experiments of the γ′($L1_2$)-precipitates in a Ni-based alloy, there is a point at which stationary coarsening occurs, which is when the volume fraction is closest to its equilibrium value. Moreover, $t_o$ *and* $\langle R(t_o)\rangle$ obtained from the NLMR method coincide with the experimental values when stationary coarsening commences, therefore they need to be included for accurate fitting of the experimental APT data.

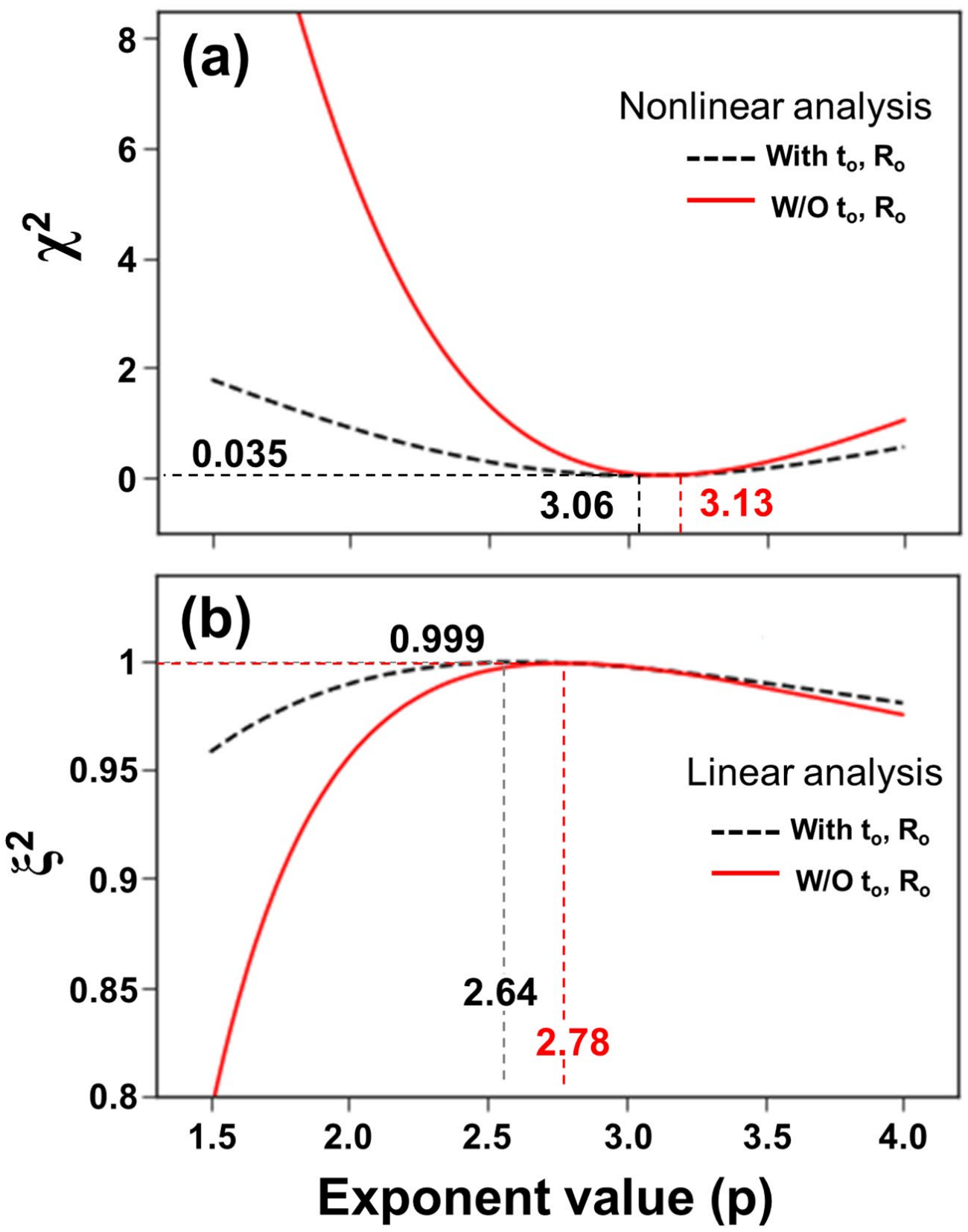

**Fig. S5.** The Chi-squared, $\chi^2$, and the coefficient of determination, $\xi^2$, as a function of the temporal exponent $p$ derived from: (a) a nonlinear multivariate regression (NLMR) analysis; and (b) linear regression analysis, respectively, with (black-dashed curve) and without (red-solid line) the variables, $t_o$ and $\langle R(t_o)\rangle$.

The difference in the best fitting value of the exponent, $p$, between nonlinear and linear analyses is due to the difference of the weighting values in the $\chi^2$- and $\xi^2$-tests. In the linear fit of $\langle R(t)\rangle^p$, the largest value of <R(t)> occurs at the longest times and it has the most impact on the goodness of fit , whereas in the nonlinear fit, the effect of the largest $\langle R(t)\rangle$ value is reduced because the square of the residuals, $(y_i - f_i)^2$, is weighted by the squared deviations for each data point, $\sigma_i^2$, Eq. S2. For this reason, although the linear regression and the coefficients of determination, $\xi^2$, are easier and simpler to calculate, it does not, however, represent the goodness of fit for curve fitting. The quantity $\langle R(t) \rangle$ is clearly not increasing *linearly* with time; therefore, the linear fit of $\langle R(t) \rangle^p$ versus time is a misleading approach for analyzing experimental data and it is also a poor predictor of how the data behave.

To summarize, determining the $p$ value using a nonlinear multivariate regression (MNLR) analysis by including the values of $t_o$ *and* $\langle R(t_o) \rangle$ is the most appropriate and accurate way of analyzing nonlinear data as opposed to choosing a value of $p$ and then plotting $\langle R(t) \rangle^p$ versus time and calculating the coefficients of determination, which has serious problems associated with it as explained above.

### *B2. Determination of the temporal exponents of the supersaturations,* $\Delta C_i^{\gamma'}(t)$.

The Chi-squared ($\chi^2$) approach used for <R(t)> is also applicable for determining the temporal exponents for the supersaturations, $\Delta C_i(t)$. The nonlinear multivariate regression (NLMR) methodology is performed by plotting the Al, Cr, and Re supersaturations versus aging time, *t*, which yields the optimized exponents, *r*, with the smallest values of Chi-squared, $\chi^2$. **Fig. S6** displays plots of $\chi^2$ versus the temporal exponent, *r,* for the solute elements, Al, Cr, and Re, in the γ(f.c.c.)-matrix and γ′($L1_2$)-precipitates, respectively. The optimized exponents, *r*, of the Al, Cr, and Re supersaturations, $\Delta C_i$, associated with the minimum values of $\chi^2$, are 2.90 ± 0.16, 2.89 ± 0.19, 2.80 ± 0.15 for the γ(f.c.c.)-matrix and 3.02 ± 0.13, 2.79 ± 0.17, 0.32 ± 0.18 for the γ′($L1_2$)-precipitate, respectively. All values of the temporal exponents (*r*) from these plots match, within error to *r* = 3, as predicted for the diffusion-limited LS model for binary alloys and the PV model for multi-component alloys. The temporal exponents, *r*, have values that are about 0.1-0.2 less than the ideal value (*r* = 3) in γ(f.c.c.)-matrix, and they are a little greater in the γ′($L1_2$)-precipitates, which is most likely because the solutes's diffusivities in γ(f.c.c.) are about three to four times faster than in the γ′($L1_2$)-precipitates, which are ordered.

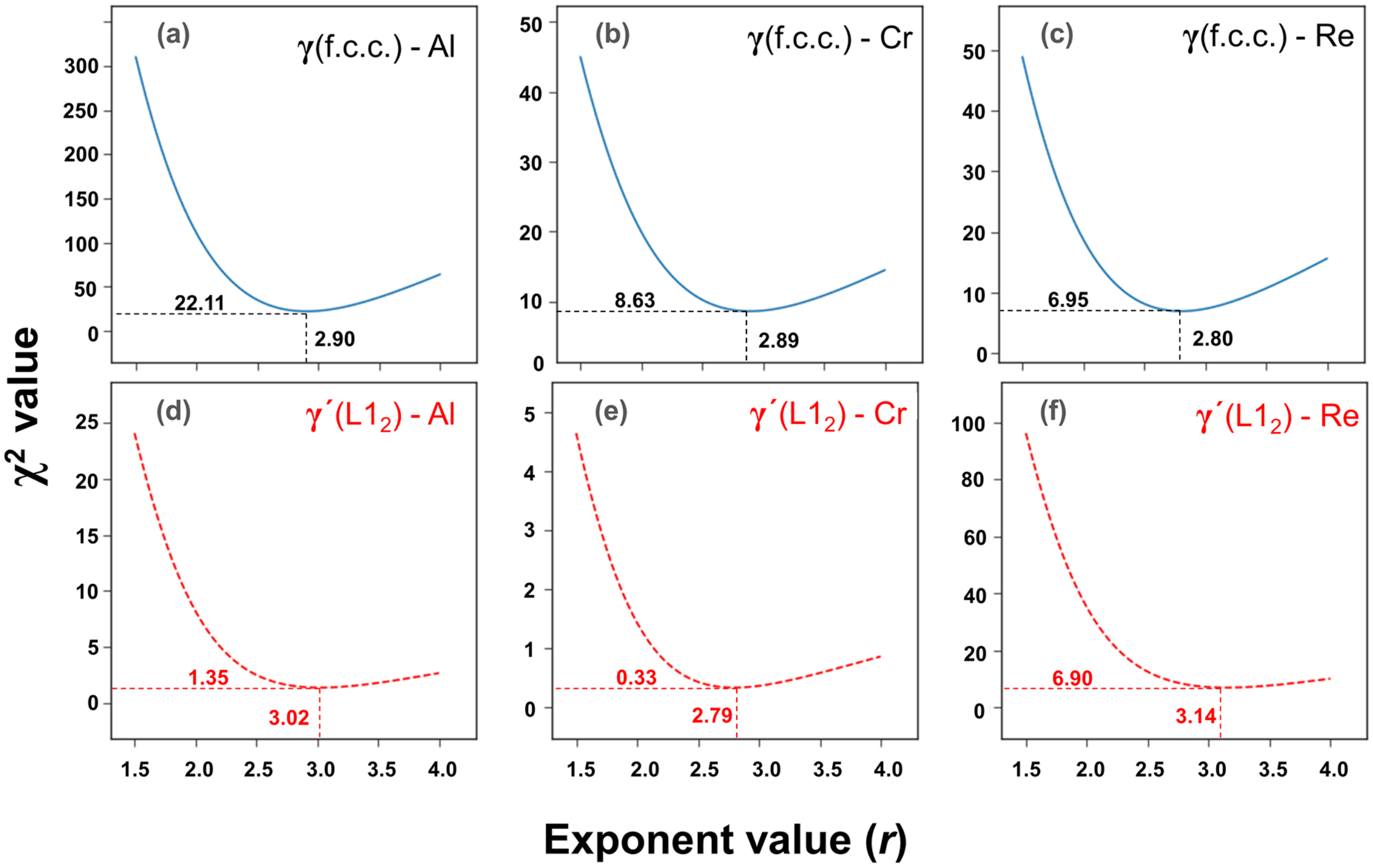

**Fig. S6**. Chi-squared, $\chi^2$, plots as a function of the temporal exponent, *r,* derived from a nonlinear multivariate nonlinear regression (NLMR) analysis of the Al, Cr, Re supersaturations: (a, b, c) in the γ(f.c.c.)-matrix; and (d, e, f) in the γ′($L1_2$)-precipitate. The minimum-$\chi^2$ and temporal exponent, *r,* values are given for each fit.

The temporal evolution of the supersaturations of Al, Cr, and Re are displayed in the **Fig. S7,** and their numerical values are listed in **Table S1**. The absolute values of $\Delta C_i$ for t ≥ 4 h were fitted using the NLMR methodology, employing the minimization of Chi-squared ($\chi^2$), **Fig. S6**, and represented by its reciprocal value (*1/r*), **Fig. S7**. The opposite direction for the relief of the Cr supersaturation for $t \leq 0.25$ h in γ′($L1_2$)-precipitates, **Fig. S7(e)**, is represented by negative values of the supersaturation. The temporal exponents are represented as the reciprocal values (*1/r*). With increasing aging times, the values of $\Delta C_i$ are decreasing continuously to achieve their equilibrium concentrations, which agrees with the PV model (red-, blue- and orange-dotted curves).

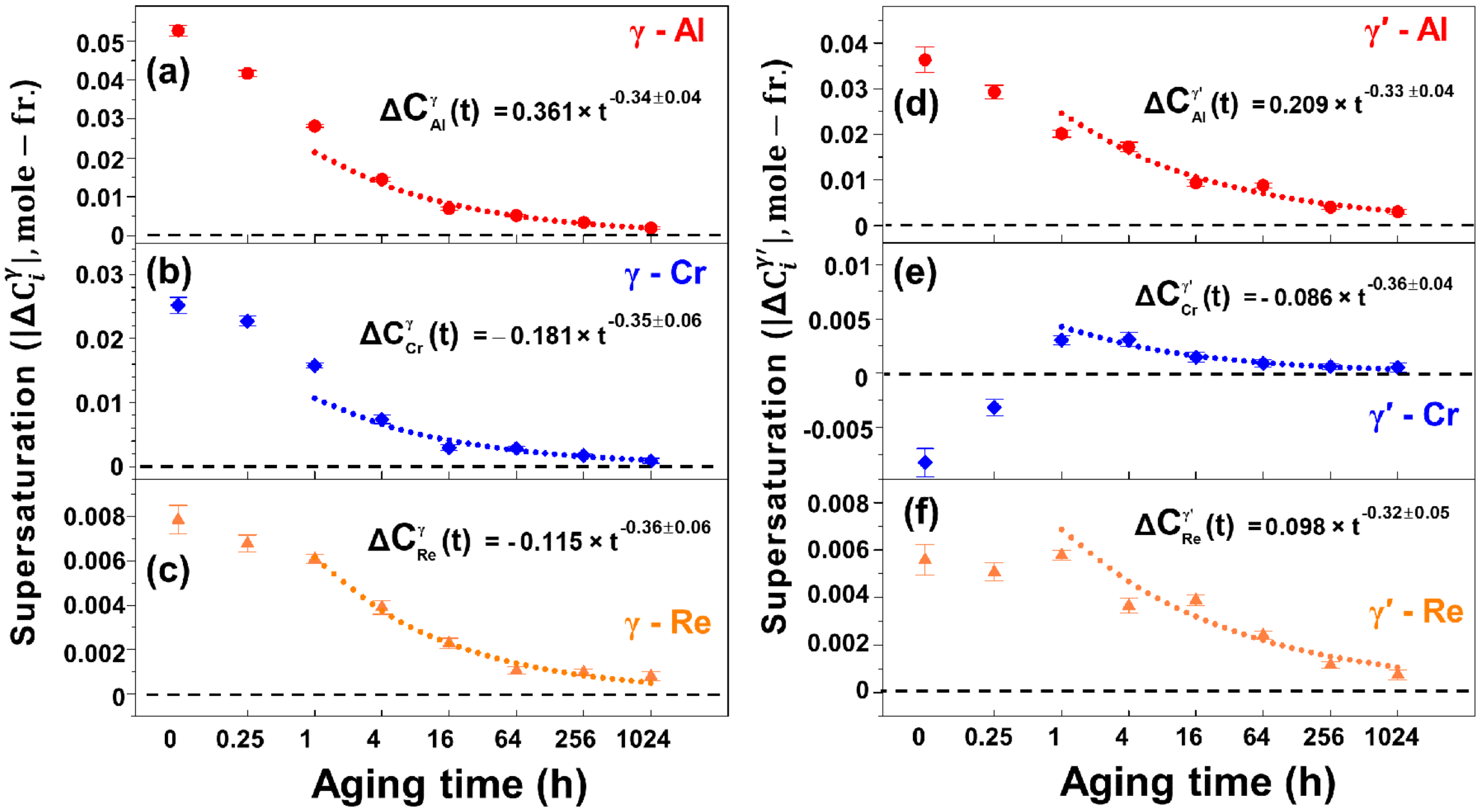

**Fig. S7**. The temporal evolution of (a, b, c) the Al, Cr, and Re supersaturations in the γ(f.c.c.)-matrix, $\Delta C_i^{\gamma}(t)$, and (d, e, f) Al, Cr, and Re supersaturations in the γ′($L1_2$)-precipitates, $\Delta C_i^{\gamma'}(t)$. The absolute values of $\Delta C_i$ for t ≥ 4 h were fitted using the NLMR methodology employing the minimization of Chi-squared ($\chi^2$) methodology. The opposite direction of the Cr supersaturation for $t \leq 0.25$ h in γ′($L1_2$)-precipitates (e) is represented as negative values of supersaturation. The temporal exponents are represented by the reciprocal value (*1/r*). Please note how small the values of the supersaturations become with increasing aging time: see **Table S1**.

**Table S1.** Temporal evolution of the supersaturations, $\left|\Delta C_i(t)\right|$, in the γ(f.c.c)-matrix and the γ′($L1_2$)-precipitate phase. The negative values of the Cr supersaturation in γ′($L1_2$)-precipitates represents the opposite direction of supersaturation for $t \leq 0.25$ h, as seen in **Fig. S7** (e) and the large scatter in the values of the supersaturation for aging times, $t \leq 1$ h, is due to the disparate diffusivities of each element in the transient regime.

| 973K | γ(f.c.c)-Matrix Supersaturations (mole-fr.) | | | |
|---|---|---|---|---|
| Time (h) | $\lvert\Delta C_{Ni}^{\gamma}\rvert$ | $\lvert\Delta C_{Al}^{\gamma}\rvert$ | $\lvert\Delta C_{Cr}^{\gamma}\rvert$ | $\lvert\Delta C_{Re}^{\gamma}\rvert$ |
| 0 | 0.0197 ± 0.0021 | 0.0527 ± 0.0016 | 0.0251 ± 0.0015 | 0.0078 ± 0.0009 |
| 0.25 | 0.0124 ± 0.0014 | 0.0417 ± 0.0010 | 0.0227 ± 0.0011 | 0.0068 ± 0.0008 |
| 1 | 0.0063 ± 0.0011 | 0.0282 ± 0.0008 | 0.0157 ± 0.0009 | 0.0061 ± 0.0007 |
| 4 | 0.0032 ± 0.0012 | 0.0145 ± 0.0009 | 0.0074 ± 0.0010 | 0.0039 ± 0.0007 |
| 16 | 0.0017 ± 0.0011 | 0.0070 ± 0.0008 | 0.0029 ± 0.0010 | 0.0023 ± 0.0006 |
| 64 | 0.0013 ± 0.0010 | 0.0052 ± 0.0008 | 0.0028 ± 0.0009 | 0.0011 ± 0.0006 |
| 256 | 0.0008 ± 0.0010 | 0.0035 ± 0.0007 | 0.0017 ± 0.0009 | 0.0010 ± 0.0005 |
| 1024 | 0.0003 ± 0.0011 | 0.0020 ± 0.0008 | 0.0009 ± 0.0006 | 0.0008 ± 0.0004 |
| | γ’($L1_2$)-precipitate Supersaturations (mole-fr.) | | | |
| Time (h) | $\lvert\Delta C_{Ni}^{\gamma'}\rvert$ | $\lvert\Delta C_{Al}^{\gamma'}\rvert$ | $\lvert\Delta C_{Cr}^{\gamma'}\rvert$ | $\lvert\Delta C_{Re}^{\gamma'}\rvert$ |
| 0 | 0.0320 ± 0.0084 | 0.0182 ± 0.0100 | -0.0082 ± 0.0059 | 0.0056 ± 0.0030 |
| 0.25 | 0.0229 ± 0.0122 | 0.0146 ± 0.0109 | -0.0031 ± 0.0063 | 0.0051 ± 0.0036 |
| 1 | 0.0128 ± 0.0052 | 0.0101 ± 0.0046 | 0.0030 ± 0.0019 | 0.0058 ± 0.0010 |
| 4 | 0.0091 ± 0.0028 | 0.0086 ±0.0025 | 0.0031 ± 0.0014 | 0.0036 ± 0.0008 |
| 16 | 0.0071 ± 0.0025 | 0.0047 ± 0.0022 | 0.0014 ± 0.0013 | 0.0038 ± 0.0007 |
| 64 | 0.0059 ± 0.0012 | 0.0044 ± 0.0010 | 0.0009 ± 0.0006 | 0.0024 ± 0.0003 |
| 256 | 0.0026 ± 0.0008 | 0.0020 ± 0.0006 | 0.0006 ± 0.0003 | 0.0012 ± 0.0002 |
| 1024 | 0.0017 ± 0.0010 | 0.0015 ± 0.0007 | 0.0005 ± 0.0004 | 0.0007 ± 0.0002 |

The linear plots of Al, Cr, Re, for comparison, are derived by fitting the supersaturations, $\Delta C_i(t)$, versus time to the power of *-1/r* ($s^{-1/r}$), **Fig. S8**. The data for $t \geq 4$ h are utilized, which is the same as for the NLMR analyses. The goodness of fit using a linear model is evaluated using the coefficients of determination, $\xi^2$. The linear plots of $\Delta C_i(t)$ versus time, *t*, to the power of *-1/r* ($s^{-1/r}$), for the largest value of the coefficient of determination, $\xi^2$, are reasonable (see inserts). These linear plots have a strong dependency on the largest value of the supersaturation at a time to the power of *-1/r* ($s^{-1/r}$). Most of the linear regression fitting lines pass through the data point at $t$ = 4 h (5th point from the left-hand side (see green circles)). The optimized values for the least-squared values of $\xi^2$ are 2.62, 2.53, 2.68 for the γ(f.c.c.)-matrix, and 3.18, 2.58, 3.58 for the γ′($L1_2$)-precipitate; these values display, however, considerably more scatter than the values obtained from the smallest Chi-squared, $\chi^2$, values in **Figs. S6** and **S7**, which is due to the sensitivity of $\xi^2$ on the largest supersaturation value at $t = 4$ h. Whereas, the nonlinear plots of the Al, Cr, and Re concentrations has an even distribution on time from 4 h to 1024 h, **Fig. S6**. As noted, the effect of the large supersaturations is weighted by the squared deviations for each data point, $\sigma_i^2$, utilizing the Chi-squared, $\chi^2$, plots. Therefore, the data plotted linearly cannot be used to identify the correct value of the temporal exponent of the supersaturations, *r*, accurately. Again, the most appropriate and accurate way of plotting the data is to plot $\Delta C_i^{\gamma}(t)$ and $\Delta C_i^{\gamma'}(t)$ versus time, *t*, and determine the temporal exponent using a nonlinear multivariate regression (NLMR) analysis.

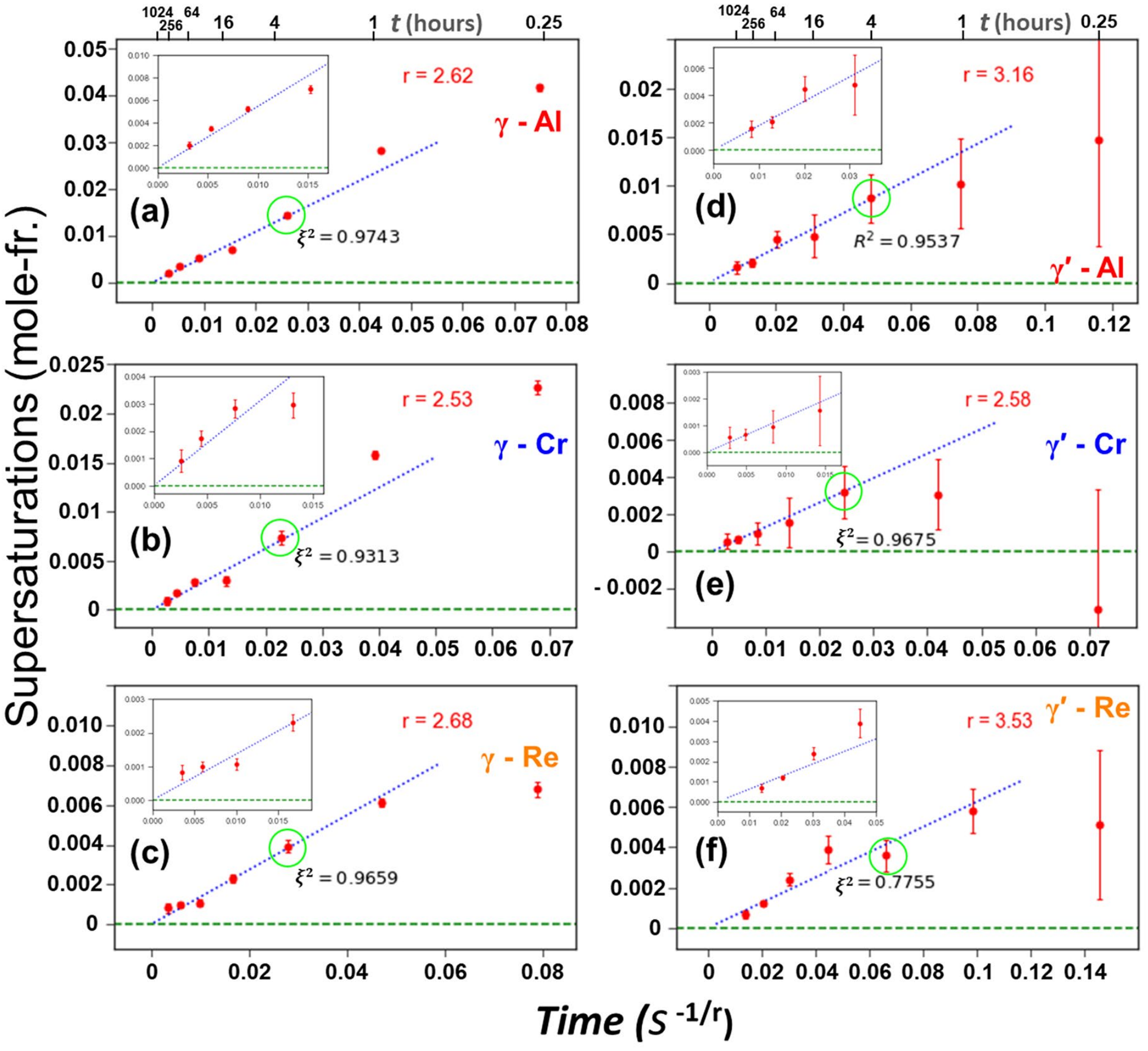

**Fig. S8**. The supersaturation, $\Delta C_i(t)$, plotted versus aging time to the power of *-1/r* ($s^{-1/r}$), for Al, Cr, Re (a, b, c) in the γ(f.c.c.)-matrix and (d, e, f) in the γ′($L1_2$)-precipitates. The maximum coefficient of determination, $\xi^2$, is given for each linear fit. It is emphasized strongly that this linear fitting procedure is not the best method for determining a temporal exponent for a supersaturation.

**Supplement C: Temporal evolution of the γ(f.c.c.)/ γ′($L1_2$)-interfacial width in a quaternary Ni-Al-Cr-Re model superalloy**

**Fig. S9** displays the concentration profiles of Ni, Al, Cr, and Re across the γ(f.c.c.)/ γ′($L1_2$)-heterophase interface for the quaternary Ni-Al-Cr-Re model superalloy after 1 h of aging. The interfacial widths, δ(t), were measured by fitting the concentration profiles to Richards' asymmetric sigmoid function [12], which is a generalized logistics function:

$$f(x, c_{\max}, c_{\min}, \theta, x_o, s) = c_{\min} + \frac{c_{\max} - c_{\min}}{(1 + v\exp(-\theta(x - x_o)))^{1/v}} \quad \text{(S6)}$$

where $c_{\min}$ is the lower asymptote, $c_{\max}$ is the upper asymptote, $\theta$ is the growth rate, $x_o$ is the shifting value of the center of a concentration profile, and $v$ is the skewness parameter. The ideal sigmoidal function is symmetric around $x_o$ with a value of $v = 1$, which is called the logistic function. Whereas when $v$ converges to zero the curve becomes the Gompertz function [13], which describes the extreme case of asymmetric growth. The diffusivities of the elements are different in the γ(f.c.c.)- and γ′($L1_2$)-phases; that is, diffusion in the ordered γ′($L1_2$)-phase is two to three orders of magnititude slower than in the disordered γ(f.c.c.)-phase. Therefore, the concentration profiles across the γ(f.c.c.)/γ′($L1_2$) heterophase can be asymmetric. Ardell [14] used a symmetric sigmoid function to measure the interfacial width; an asymmetric concentration profile cannot, however, be fitted accurately with this function. The concentration profiles are initially asymmetric. This problem can be corrected by employing a skewness parameter, $v$, in Eqn. S6. The interfacial widths, δ(t), are measured from the horizontal distances between the $10^{th}$ and $90^{th}$ percentile values of the concentration differences between the γ(f.c.c.)- and γ′($L1_2$)-phases. The five parameter Richards's function produces a more accurate fit to the experimental data with smaller Chi-Squared, $\chi^2$ -values. This asymmetric tendency is more noticeable at earlier aging times. At a t =1 h, all the concentrations at the γ(f.c.c.)/γ′($L1_2$)-heterophase interface are skewed toward the left-hand side of the γ(f.c.c.) phase with $v$ being <1. The Re concentration profile has the smallest skewness value, $v = 0.01$. This skewness disappears at longer aging times, and the concentration profiles become symmetrical with $v = 1$. The width of each interface is represented by two dashed vertical green lines at $x$-positions that coincide with the $10^{th}$ and $90^{th}$ percentiles of the concentration difference between the γ(f.c.c.)- and γ′($L1_2$)-phases. The shifting value of the center, $x_o$, is located close to the left-hand side of the interfacial region because of the smaller value of the skewness parameter, which is < 1. Another important feature is that each element has a different interfacial width. For example at an aging time of 1 h, Cr has the largest value, 1.88 ± 0.15 nm, whereas Re has the smallest value, 1.68 ± 0.17 nm, which is a reflection of their differerent diffusivities in Ni. At an earlier aging time, t ≤1 h, the partitioning of Re between the γ(f.c.c.)- and γ′($L1_2$)-phases has the smallest value. Hence, the γ(f.c.c.)/ γ′($L1_2$) heterophase interface for Re is in a transient state at t ≤1 h.

The measured interfacial widths, $\delta_i(t)$, between the γ(f.c.c.) and γ′($L1_2$) phases are normalized by the mean radius, $\langle R(t) \rangle$, **Fig. S10**. The resulting normalized widths, $\frac{\delta(t)}{\langle R(t) \rangle}$, decrease unambiguously with increasing $\langle R(t) \rangle$ by a factor of more than 10 as the time varies from 0.25 to 1024 h for all the elements. A NLMR analysis of the normalized interfacial widths was performed employing a decaying power-law function:

$$\frac{\delta(t)}{\langle R(t) \rangle} = \alpha \langle R(t) \rangle^{\beta} \tag{S6}$$

where $\alpha$ is the rate constant, and $\beta$ is the temporal exponent. All the temporal exponents , $\beta$ , are negative and dimensionless: -1.17 ± 0.02, -1.07 ± 0.01, -0.97 ± 0.01, and -0.82 ± 0.01 for Ni, Al, Cr, and Re, respectively. The absolute values of the interfacial widths of Ni and Al decrease, whereas those of Cr and Re increase with increasing time and mean radius. While all the normalized widths, $\frac{\delta(t)}{\langle R(t)\rangle}$, decrease, however, with increasing time to first order as $\langle R(t)\rangle^{-1}$. The trans-interface-diffusion-controlled (TIDC) model [15-17] posits that the mean precipitate radius, $\langle R(t)\rangle$, increases with increasing aging time, t, according to the equation $\langle R(t)\rangle^p = Kt$, where $p$ is an exponent ($2 \le p \le 3$) related to the width of the interface. Ardell and Ozolins [15] obtained the optimized $p$-value using the precipitate size distribution (PSD) and an optimized coefficient of determination, $\xi^2$, which is incorrect because of the nonlinear behavior of the temporal evolution of $\langle R(t)\rangle$, **eqn. S3**. The TIDC model also assumes that the fluxes through the interface are proportional to the differences between the concentrations or chemical potentials of each element in the interfacial profiles. Based on this assumption, Ardell and Ozolins [15] made the ansatz that the interfacial width, $\delta_i(t)$, increases with increasing aging time, $t$, as $\langle R(t)\rangle^p = Kt$, where $p = m + 2$: $m$ is the exponent's value for the temporal evolution of $\delta_i(t)$. Our APT measurements are, however, inconsistent with the TIDC model's assumptions. That is, we observe an increasing mean radius, $\langle R(t)\rangle$, and concomitantly decreasing normalized interfacial widths, $\frac{\delta(t)}{\langle R(t)\rangle}$, with increasing aging time for all the Ni-based alloys we have studied to date: binary[18], ternary[5, 6, 19-21], quaternary[22-25], quinary[26, 27], and sexinary [28].

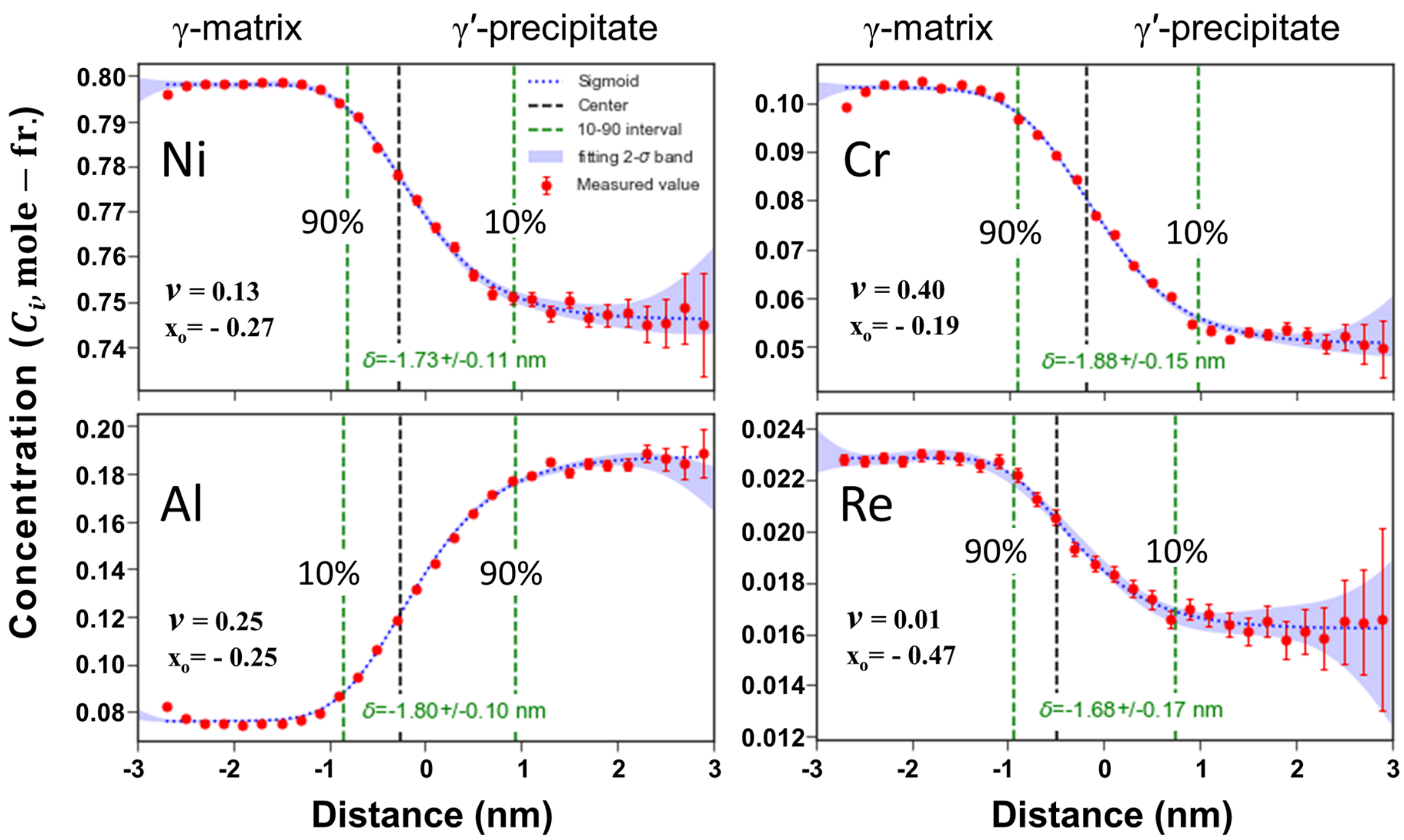

**Fig. S9**. The concentration profiles of Ni, Al, Cr, and Re across the γ(f.c.c.)/γ′($L1_2$)-heterophase interface for the quaternary Ni-Al-Cr-Re model superalloy after 1 h at 973 K. Positive distances are into the γ′($L1_2$) precipitates, while negative distances are into the γ(f.c.c.) matrix. The interfacial widths, δ(t), were measured by fitting the concentration profiles to Richards' asymmetric sigmoid function using the horizontal distances between the $10^{th}$ and $90^{th}$ percentiles of the concentration differences between the γ(f.c.c.) and γ′($L1_2$) phases. The parameter, $\nu$ , represents the skewness of the sigmoid function, **Eqn. (S5).**

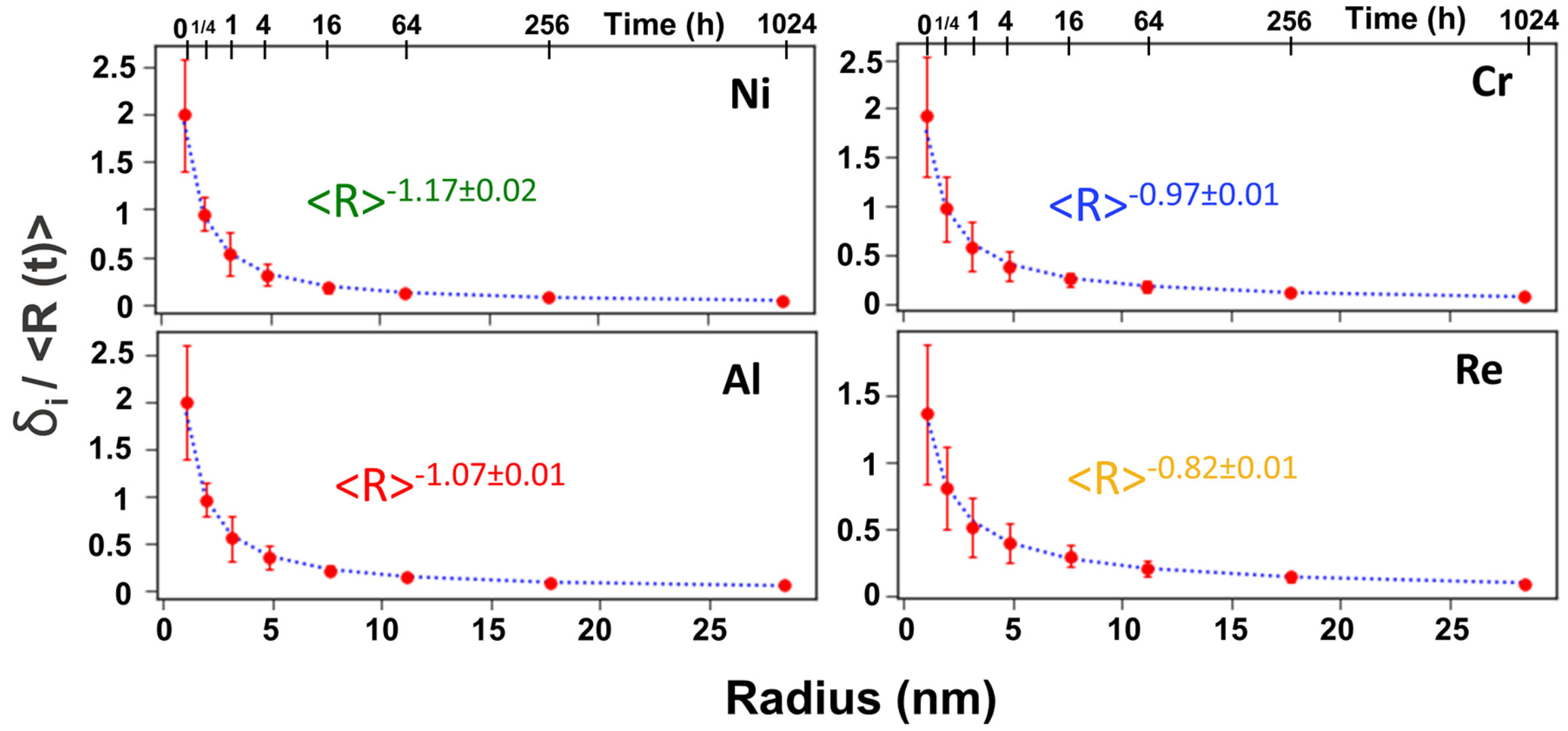

**Fig. S10**. Interfacial width, $\delta_i(t)$, between the γ(f.c.c.) and γ′($L1_2$) phases normalized by the mean radius, $\langle R(t)\rangle$, versus <R(t)>. The time scale in hours (h) is indicated on the upper abscissa of the plots and the experimental data points have plus or minus two standard deviations (2σ). It is clear from these experimental results that this ratio decreases with increasing <R(t)>.